\documentclass[reprint,amsmath,amssymb,aps]{revtex4-2}
\usepackage{graphicx}
\usepackage{placeins}
\usepackage{booktabs}
\usepackage{siunitx} 
\usepackage{dcolumn}
\usepackage{bm}

\usepackage{epstopdf}

\usepackage{xcolor}
\begin{document}
\title{Ro-vibrational quenching calculations of C$_2^-$  in collision with H$_2$}
\author{Kousik Giri}
\affiliation{Department of Computational Sciences, Central University of Punjab, Bathinda, Punjab 151401, India}
\author{Barry Mant}
 \affiliation{Instit{\"u}t f{\"u}r Ionenphysik und Angewandte Physik, 
Universit{\"a}t Innsbruck, Technikerstr. 25, 6020, Innsbruck, Austria}
\author{Franco A. Gianturco}
\affiliation{Instit{\"u}t  f{\"u}r Ionenphysik und Angewandte Physik, 
Universit{\"a}t Innsbruck, Technikerstr. 25, 6020, Innsbruck, Austria}
\email{francesco.gianturco@uibk.ac.at}
\author{Jan Franz}
\affiliation{Institute of Physics and Applied Computer Science, 
 Faculty of Applied Physics and Mathematics, Gda\'nsk University of 
Technology, ul. Narutowicza 11/12, 80-233 Gda\'nsk, Poland}
\author{Rupayan Biswas}
\affiliation{School of Chemical Sciences, National Institute of Science Education and Research (NISER) Bhubaneswar,\\
An OCC of Homi Bhabha National Institute,  Khurda, Odisha 752050,India}
\author{Upakarasamy  Lourderaj}
\affiliation{School of Chemical Sciences, National
Institute of Science Education and Research (NISER) Bhubaneswar,\\
An OCC of Homi Bhabha National Institute,  Khurda, Odisha 752050,India}
\author{Narayanasami  Sathyamurthy}
\affiliation{Indian Institute of Science Education and Research Mohali, SAS Nagar, Punjab 140306, India}
\author{Roland Wester}
\affiliation{Instit{\"u}t  f{\"u}r Ionenphysik und Angewandte Physik, 
Universit{\"a}t Innsbruck, Technikerstr. 25, 6020, Innsbruck, Austria}

\date{\today}

\begin{abstract}
The molecular anion C$_2^-$ has been of interest in the last few years as a candidate for laser cooling due to its electronic structure and
favourable branching ratios to the ground electronic and vibrational state. Molecular hydrogen has been used by the Wester group in Innsbruck as a buffer gas to cool 
the molecule's internal ro-vibrational motion. In the present work, we generate a new, five dimensional (5D) interaction potential for the system  by considering the H$_2$ as a rigid rotor and the C$_2^-$ as a rotating-vibrating diatomic molecule. We thereafter calculate the cross sections and rate coefficients for ro-vibrational inelastic collisions of
C$_2^-$ with both para- and ortho-H$_2$ on this new 5D \textit{ab initio} potential energy surface using quantum scattering theory for the dynamics. 
The rates for vibrational quenching are obtained over the range of temperatures which covers the single value measured by the experiments. A comparison is also made with the earlier results using a simpler 3D interaction potential. Furthermore, para-H$_2$ is found  to be more efficient than ortho-H$_2$ (with or without undergoing rotational excitation) in cooling C$_2^-$. The rate coefficients for cooling the anions has been computed by appropriately weighting the ortho- and para-H$_2$ and compared with the available experimental result at 20 K. When the vibrational de-excitation rate coefficients are  taken to be the ones not causing any concurrent rotational excitations in the final C$_2^-$ anions, the properly  averaged results are found to get smaller and to become very close to the experimental measurements. The implications of these new results for laser cooling of C$_2^-$ are analyzed and discussed.
\end{abstract}

\maketitle

\section{Introduction}
\label{sec:intro}

Laser cooling of molecules is currently an active area of research \cite{Tarbutt2018:cp} since, with the direct preparation of ultracold molecular
ensembles in magneto-optical traps \cite{Barry2014:n}, numerous experiments on molecular quantum control, precision spectroscopy \cite{13LoCoGr}, and ultracold chemistry \cite{19DoEbKo} have become accessible. For atoms, laser cooling of 
neutral and charged species has been a workhorse for many years, while
for molecules no charged molecular species has yet been successfully laser cooled.  The main reason for it is that , as it happens, molecular ions with bound excited electronic states below the first fragmentation threshold which can be excited with suitable narrow-band lasers are hard to come by. Furthermore, a near optimal 
Franck-Condon overlap of the vibrational wave functions of the ground and excited electronic states is required to make the closed optical cycles feasible. 

As a matter of fact, the diatomic carbon molecular anion has been identified as an interesting exception \cite{15YzHaGe}, as it possesses several bound excited
electronic states below the photodetachment threshold. Furthermore, the 
electronic states $A^2\Pi_u$ and $B^2\Sigma_u^+$ are known to have reasonably high Franck-Condon overlap factors with the $X ^2\Sigma^+_g$ ground state for the
transitions between their lowest vibrational levels $\nu'=0 \rightarrow \nu''=0$ \cite{03ShXiBe, 24RSS, 16ShLiMe.c2m}. Simulations of laser cooling
using the $B^2\Sigma_u^+$ \cite{15YzHaGe} and $A^2\Pi_u$ \cite{17FeGeDo} states have shown that C$_2^-$ can, in principle, be cooled
efficiently to millikelvin temperatures using Doppler or Sisyphus cooling in Paul or Penning traps. If laser cooling of C$_2^-$ anions were to be realised, it would open up the
possibility of sympathetically cooling other anions \cite{17FeGeDo} or even antiprotons \cite{18GeFeDo}, since the efficient
production of antihydrogen atoms is  being investigated for tests of fundamental physics such as CPT invariance \cite{17Ahxxxx} 
and the weak equivalence principle \cite{12PeSaxx}.

The diatomic  molecular anion C$_2^-$ has been a model system for decades, attracting a great deal of experimental 
\cite{68HeLaxx.c2m,69MiMaxx.c2m,71.Frxxxx.c2m,72LiPaxx.c2m,80JoMeKo.c2m,82LeNaMi.c2m,85MeHeSc.c2m,88ReLiDi.c2m,91ErLixx.c2m,
92RoZaxx.c2m,95BeZhYu.c2m,98PeBrAn.c2m,03BrWeDa.c2m,17Naxxxx.c2m,14EnLaHa.c2m} and theoretical 
\cite{74Baxxxx.c2m,79ZePeBu.c2m,80DuLixx.c2m,84RoWexx.c2m,87NiSixx.c2m,92WaBaxx.c2m,06SeSpxx.c2m,16ShLiMe.c2m,19KaLoLi.c2m,19GuJaKr} interest. 
Its electronically excited states \cite{87NiSixx.c2m} are still bound states, a feature that is unusual for an anion and which is a consequence of the high electron affinity (EA)  
of neutral C$_2$  (around 3.3 eV \cite{80JoMeKo.c2m,91ErLixx.c2m}) in combination with the open shell character of the electronic
configuration of carbon dimer. In its ground electronic state ($X ^2\Sigma^+_g$) the molecule has only evenly numbered rotational states due to the nuclear 
statistics of the $^{12}$C$_2^-$ molecule with zero spin nuclei, while in the excited $B^2\Sigma_u^+$ state, only odd-numbered rotational states exist.

Furthermore, the C$_2^-$ anion could also be present in astrophysical environments since the neutral C$_2$ is abundant in interstellar space \cite{95LaShFe.c2m} and in comet tails \cite{90LaShDa.c2m} and is a common component of 
carbon stars \cite{77SoLuxx.c2m,86LaGuEr.c2m}. The large EA of C$_2$ and strong electronic absorption bands of 
C$_2^-$ \cite{72LiPaxx.c2m} could make the anion  detectable in space \cite{80VaSwxx.c2m} but as yet, also due to its lack of a permanent dipole moment, no conclusive evidence of its presence has been found \cite{72FaJoxx.c2m,82Waxxxx.c2m,05CiHoKa.c2m}. Its most abundant isotopologue $^{12}$C$_2^-$
is an homonuclear diatomic molecule which does not exhibit a pure ro-vibrational spectrum, thus making difficult  detection in emission of this anion. 

Laser cooling of C$_2^-$ would ideally start with ions initially cooled to around 10 K, for example by helium buffer gas cooling in a
cryogenic ion trap \cite{Wester2009:jpb,19GiGoMa.c2hm}. Processes used to generate C$_2^-$ involve applying an electric discharge to a 
mixture of C$_2$H$_2$ and CO$_2$ in a carrier gas \cite{03BrWeDa.c2m,19HiGeOs} which may form the anion in its excited vibrational states.
Besides cooling the translational motion, the buffer gas is then also required to cool the internal degrees of freedom via inelastic collisions.
Furthermore, the buffer gas may be a useful tool to quench excited vibrational levels when they get populated during laser cooling due to
the non-diagonal Franck-Condon factors. This could circumvent the need for additional re-pumping lasers. In a similar scheme, 
rotational buffer gas cooling was performed during sympathetic translational cooling of MgH$^+$ \cite{14HaVeKl}.

In a recent paper we reported cross sections and rate coefficients for C$_2^-$-He rotationally inelastic collisions, treating the anion 
as a rigid rotor \cite{20MaGiGo}. The computed rates for rotational excitation and quenching were found to be in line with those 
for similar ionic species interacting with helium \cite{19GiGoMa.c2hm}. Simulations of cooling rotational motion at typical helium 
pressures in ion traps showed that thermalisation to Boltzmann populations occurred within a fraction of a second. We have extended 
this work by also modelling the rotational cooling of C$_2^-$ with neon and argon atomic partners \cite{20aMaGiWe}. It was found that thermalization 
times of C$_2^-$ with He and Ne were fairly similar but cooling was significantly faster with Ar. This is obviously due to the increased interaction
strength between C$_2^-$ and the largest atom  in the series He $<$ Ne $<$ Ar. Then we have additionally investigated rovibrational inelastic rate coefficients of the present title anion  with He, Ne  and Ar \cite{20MaGiWe}, showing that such energy-exchange processes carry fairly small probabilities for He and Ne, while the Ar partner produces the largest inelastic rates along the series.

In a recent experimental investigation of the rovibrational quenching processes in collision with H$_2$ \cite{23NWL}, we have compared the measured experimental rate coefficient value with quantum calculations in which the molecular hydrogen partner was treated as a structureless object, thereby reducing the number of dimensions of the interaction potential energy surface (PES) to three. Since the agreement of the calculations with experiment turned out to be inadequate, we have now extended the number of dimensions of the required PES to five, as mentioned earlier and as discussed in the next Section. Therefore, we shall present in this work the new features of the interaction forces, together with the new quantum calculations of the relevant inelastic cross sections and inelastic rate coefficients. The concurrent rotational excitation processes involving the H$_2$ will also be presented and discussed.

As C$_2^-$ has no oscillating dipole moment, the vibrational levels are long lived with the ground electronic
state's $\nu = 2$ levels persisting for over five seconds \cite{98PeBrAn.c2m} and so collisions are the only viable efficient means of quenching these
states. The rate coefficients for C$_2^-$ vibrational quenching with H$_2$ may also prove useful in future astrophysical studies, 
should the anion be detected in an interstellar environment where excited vibrational states are important for observation, as would be the case  in the
circumstellar envelope around carbon rich stars where molecular hydrogen is also abundant.

The paper is organised as follows. In the next section we discuss the calculations  carried out for the new \textit{ab initio} multidimensional potential energy surface for the
 C$_2^-$ in its ground electronic ($^2 \Sigma_g^+$) state interacting with the H$_2$ molecule, also in its ground electronic state.  In that Section
 we provide details of the \textit{ab initio}  5D  PES and describe the fitting via a neural network generated functional form to the raw original points. This section also contains details of the vibrationally averaged matrix elements required for 
scattering calculations. The  close-coupled scattering calculations which we have employed in this study are described in Section \ref{sec:scat}. Cross sections 
and corresponding rate coefficients for rotationally and vibrationally inelastic collisions are presented in Section \ref{sec:results}. We summarize our
present findings in Section \ref{sec:conc}, where we also report our conclusions.

\section{C$_2^-$ ($X^2 \Sigma_g^+$) interacting with the H$_2$ molecular partner: details of the 5D Potential Energy Surface} 
\label{sec:PES}

As  discussed recently by Gulania \textit{et al.} \cite{19GuJaKr}, the electronic structure of the C$_2$ molecule is notoriously 
difficult to calculate accurately due to many low-lying electronic states giving rise to a multi-reference character of the  
\textit{ab initio} description of its ground electronic state. For the C$_2^-$ anion considered here, the situation is not so severe but the presence of a close-lying $A^2\Pi_u$ state (4000 cm$^{-1}$ above the ground ($^2 \Sigma_g^+$) state) still makes electronic structure calculations challenging.

\begin{figure}
\centering
\includegraphics[width=\linewidth]{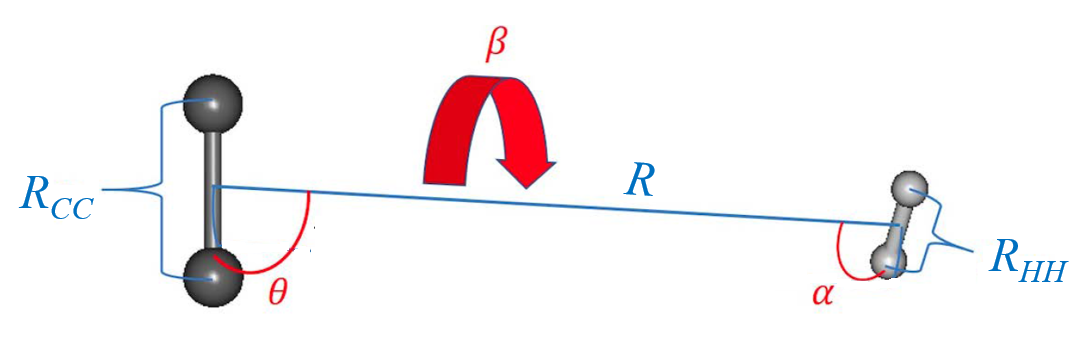}
        \caption{Graphical representation of the five active coordinates employed to generate the 5D PES of the present system. The distance between the centre of mass of C$_2^-$ and that of H$_2$  is the radial distance $R$. The two bond distances are: $R_{CC}$ and $R_{HH}$. The three angles are defined in the main text. See main text for further details.}
        \label{fig: Fig1}
\end{figure}
The interaction energies between C$_2^-$ in its ground  electronic state and  the H$_2$ molecular partner in its ground electronic state were calculated to map the full 5D PES as a function of the five independent variables pictorially shown in Figure \ref{fig: Fig1}. We have used \textit{ab initio} methods as implemented in the MOLPRO suite of codes, Version 2022.2  \cite{MOLPRO,MOLPRO_brief}. Geometries of the C$_2^-$ - H$_2$ system were defined on a Jacobi grid as illustrated  
in Figure 1. In the calculations, the value of $R$ ranges between 2.6 \AA\ and 25.0 \AA\ (the choice of grid points varies with the angles and the $R_{\textnormal{CC}}$ bond length). The $\theta$, $\beta$ and $\alpha$ angles are defined between $R$ and 
the two internuclear bond axes $R_{\textnormal{CC}}$ and $R_{\textnormal{HH}}$ as shown in the graphic reported by Figure \ref{fig: Fig1}. For the angle $\theta$, values at intervals of 10$\ensuremath{^\circ}$  from 0  to 90\ensuremath{^\circ}\ were used. For the $\beta$ angle four values at intervals of 30$\ensuremath{^\circ}$ between 0  and 90$\ensuremath{^\circ}$ were used. Finally, for the  angle $\alpha$, we employed seven values between 0  and 180$\ensuremath{^\circ}$ at intervals of 30$\ensuremath{^\circ}$. The values of the C-C bond length 
were varied  between 1.18 and 1.50 \AA\ (1.18, 1.22, 1.24, 1.2689, 1.28, 1.30, 1.35, 1.40, 1.45, 1.50 \AA), including the equilibrium value of  $R_\textnormal{CC}$$_{eq}$ = 1.2689 \AA. This is sufficient to cover the vibrational
levels of C$_2^-$ of interest in the present study. 

More specifically, we shall be dealing mainly with the transition involving the $\nu = 1$ to $\nu= 0$ vibrational levels of C$_2^-$, with an energy difference calculated by us to be about 1754 cm$^{-1}$, a value not far from the experimental gap of 1757 cm$^{-1}$ as discussed in our earlier work \cite{23NWL}. Our computed rotational constant was $B_{\textnormal{e}}$ = 1.729 cm$^{-1}$ (see a comparison with other data in our earlier work in\cite{23NWL}).
Interaction energies between C$_2^-$ and the H$_2$ partner were determined by 
subtracting the asymptotic energies for each  $R_{\textnormal{CC}}$ value. Because the energy separation between the lower vibrational levels of the hydrogen molecular partner is two to three times larger than those of the anion, we expect that its vibrational excitation probabilities at the low temperatures of interest in this study would be much smaller than those for the anion,  especially for the temperatures of interest in this work ($T$ $\le$ 100 K). Furthermore, the energy spacing between  the $j_2 = 0$ and $j_2 = 2$ levels of $p$-hydrogen is around 500 K.  Hence, we have disregarded the vibrational coordinate ($r_2$ = $R_{\textnormal{HH}}$) of the neutral partner indicated in Figure 1. On the other hand, we have included the dynamics of concurrent rotational excitations of the H$_2$ partner in some of our calculations. 

The potential energy values for the system were calculated using the restricted coupled clusters singles, doubles and perturbative triples (RCCSD(T)-F12) method \cite{85WeKnxx,85KnWexx}, including relativistic corrections \cite{11ShKnWe.LM}. An aug-cc-pVQZ basis set \cite{92KeDuHa} was used on each carbon centre and 
an aug-cc-pV5Z basis set for the H atoms.

\subsection{The Artificial Neural Network fitting of the 5D raw points} 
\label{sec:NN}
The 5D \textit{ab initio} data were then fitted using the artificial neural network (ANN) method. For an extended documentation on the method see:\cite{behler2021four, meuwly2023neural, dhzhang2023accurate}. The data set consisted of 76,474 points (the number of original raw points computed in the five independent variables described earlier was a bit larger: 76,634. It was slightly reduced to improve the smoothness of their representation). That set of points covered an  energy range from -723.81 cm$^{-1}$ to 10997.2 cm$^{-1}$. The Levenberg-Marquardt algorithm along with the Bayesian Regularization method was used to adjust the neural network parameters.   The activation functions were chosen to be modified sigmoid functions given as:
\begin{equation}
\sigma(a) = 2/(1+\exp(-2a) ) - 1 
\end{equation}
The target mean-squared-error (MSE) was set at 0.5 cm$^{-1}$ and the ANN was trained for a maximum of 4000 epochs. The data set was partitioned into training and test data sets. Different fits  were obtained using a two layer (40, 40) network, where the numbers indicate the chosen number of nodes in each layer. We then  varied the training/test ratio from 60/40 to 90/10. It was found that the 90/10 ratio gave the best root-mean-squared-deviation (RMSD) of 36.36 cm$^{-1}$ for the test data. Further increase in the number of nodes did not improve the quality of the fit. To improve the fit, different fits with three layered networks were attempted for a 90/10 training/test ratio. The best fit (fit 9) was obtained by a (50, 50, 50) network that resulted in an  RMSD of 0.70 cm$^{-1}$ for the training set and an RMSD of 9.05 cm$^{-1}$ for the test data. A summary of the ANN fits is given in Table \ref{tab:ANN}. The residuals for the best fit are plotted in a Figure provided by the Supplementary Information. The maximum deviation was found to be 693.58 cm$^{-1}$ that occurred for an energy of 7618.85 cm$^{-1}$. It should be noted that for the energy range considered in the quantum dynamical study reported below, this  deviation is not consequential. The ANN PES thus obtained was found to be smooth without over-fitting issues as illustrated in another Figure also provided in the Supplementary Information Section.

\begin{table}[h]
\centering
\caption{Summary of the ANN fits obtained for the C$_2^-$/H$_2$ system}
\begin{tabular}{ccccc}
\toprule
 & Training & & \multicolumn{2}{c}{RMSD (cm$^{-1}$)} \\
\cline{4-5}
 Fit  &  Data (\%) & NN & Test & Train\\
\midrule
 1 & 60 & 40,40 & 63.49 & 2.10\\
 2 & 70 & 40,40 & 411.8 & 2.56\\
 3 & 80 & 40,40 & 298.94 & 1.55 \\
 4 & 90 & 40,40 & 36.36 & 2.88 \\
 5 & 90 & 50,50 & 71.15 & 3.66\\
 6 & 90 & 60,60 & 154.23 & 0.783\\
 7 & 90 & 50,40,40 & 28.72 & 0.70  \\
 8 & 90 & 50,50,40 & 21.51 & 0.70  \\
 9 & 90 & 50,50,50 & 9.05 & 0.70  \\
\bottomrule 
\end{tabular}
\label{tab:ANN}
\end{table}

After having presented in the earlier part of this Subsection the details of the ANN fitting of the initial raw points generated for the 5D PES discussed earlier, it is interesting to see how our new results compare with the earlier evaluation in three dimensions (3D) \cite{23NWL}. To this end, we show in Figure 2, in its four panels, a reduction to 3D of the present 5D PES with a further comparison with the same data from our earlier paper \cite{23NWL}. We indicate in those four panels two different bond lengths of the vibrating C$_2^-$ for two different angular orientations of the H$_2$ partner, treated in both cases as a point-like structure. It is clear from the comparison in the figure that the present results reproduce very well the earlier calculations, a reassuring feature which confirms the quality of the present study treating the interaction forces in five dimensions.

\begin{figure}
\centering
\includegraphics[width=\linewidth]{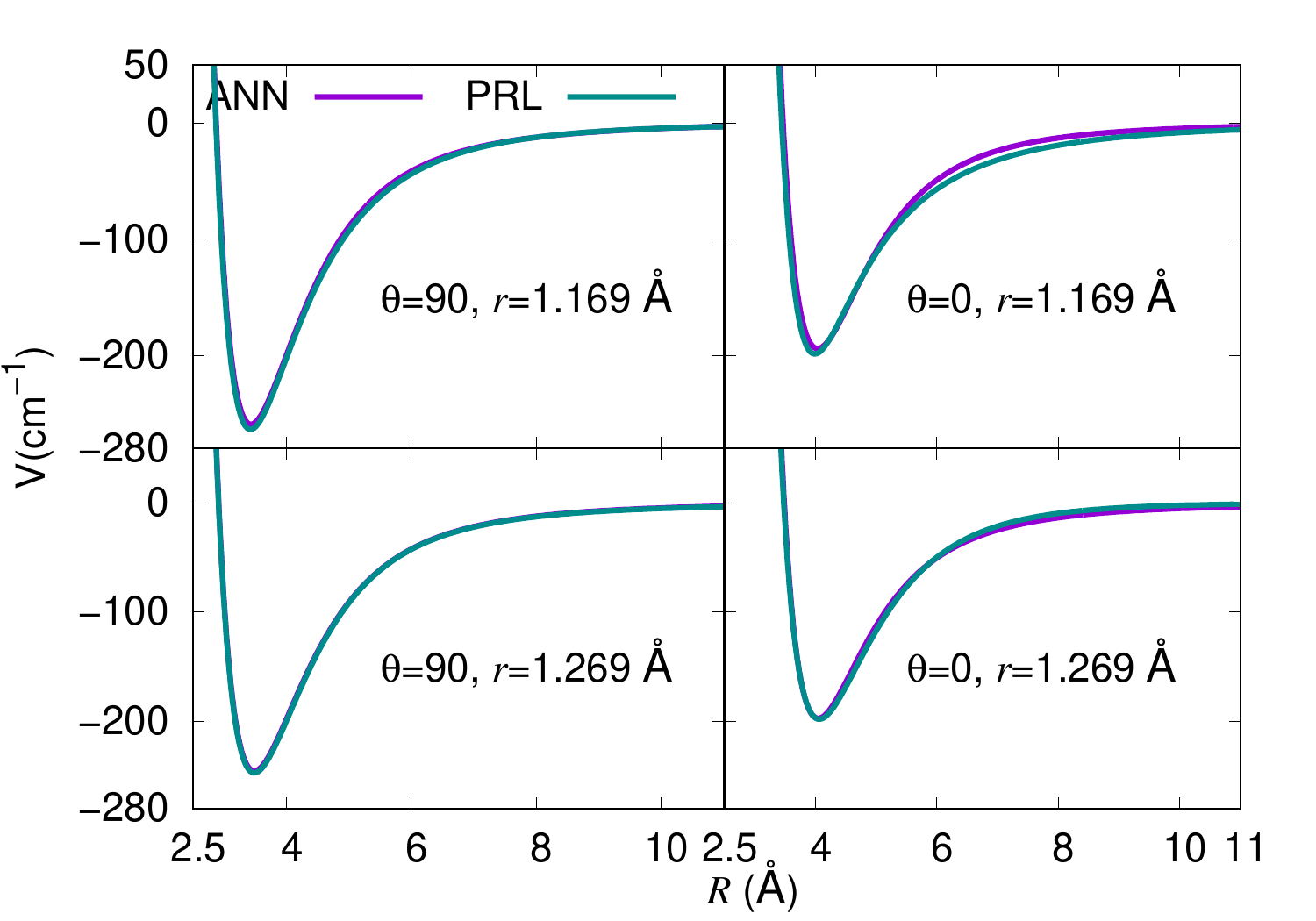}
        \caption{Graphical presentation of a comparison between the new calculations of the present work (the ANN results) and the earlier 3D results reported by us in reference \cite{23NWL}. Here they are labelled as  the PRL (Phys.Rev.Lett.)  lines. See the main text for further details.}
        \label{fig: Fig4}
\end{figure}

To further verify the  quality of the  ANN fit, we have  checked in detail the range of the C-C bond length for which the ANN expansion had sufficient angular and internuclear distance points. We actually found it to be  expedient to constrain the ranges of the original configurations  to avoid nonphysical features in the resulting 5D interaction for $R_{\textnormal{CC}}$ distances larger than 1.4 \AA\ . The effects of such numerical limitations to help in data smoothness are reported in three sets of panels of Figure 3, which show how we produce in the end  smooth-behaving angular cuts for a variety of angles in the fuller 5D interaction potential.

\begin{figure}
\centering

\label{fig3a}
	\includegraphics[width=1.0\linewidth,angle=+0.0]{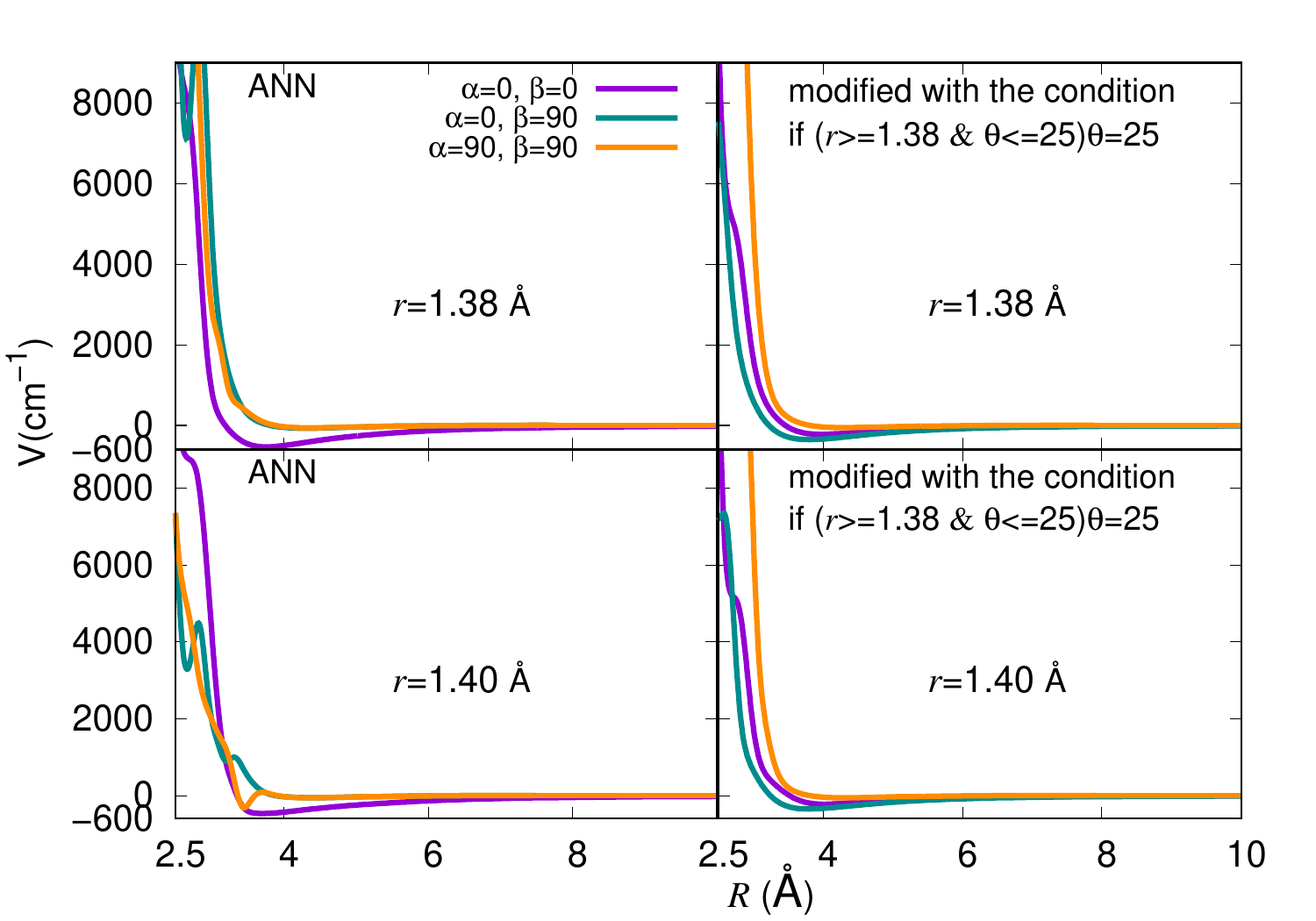} 

\label{fig3b}
	\includegraphics[width=1.0\linewidth,angle=+0.0]{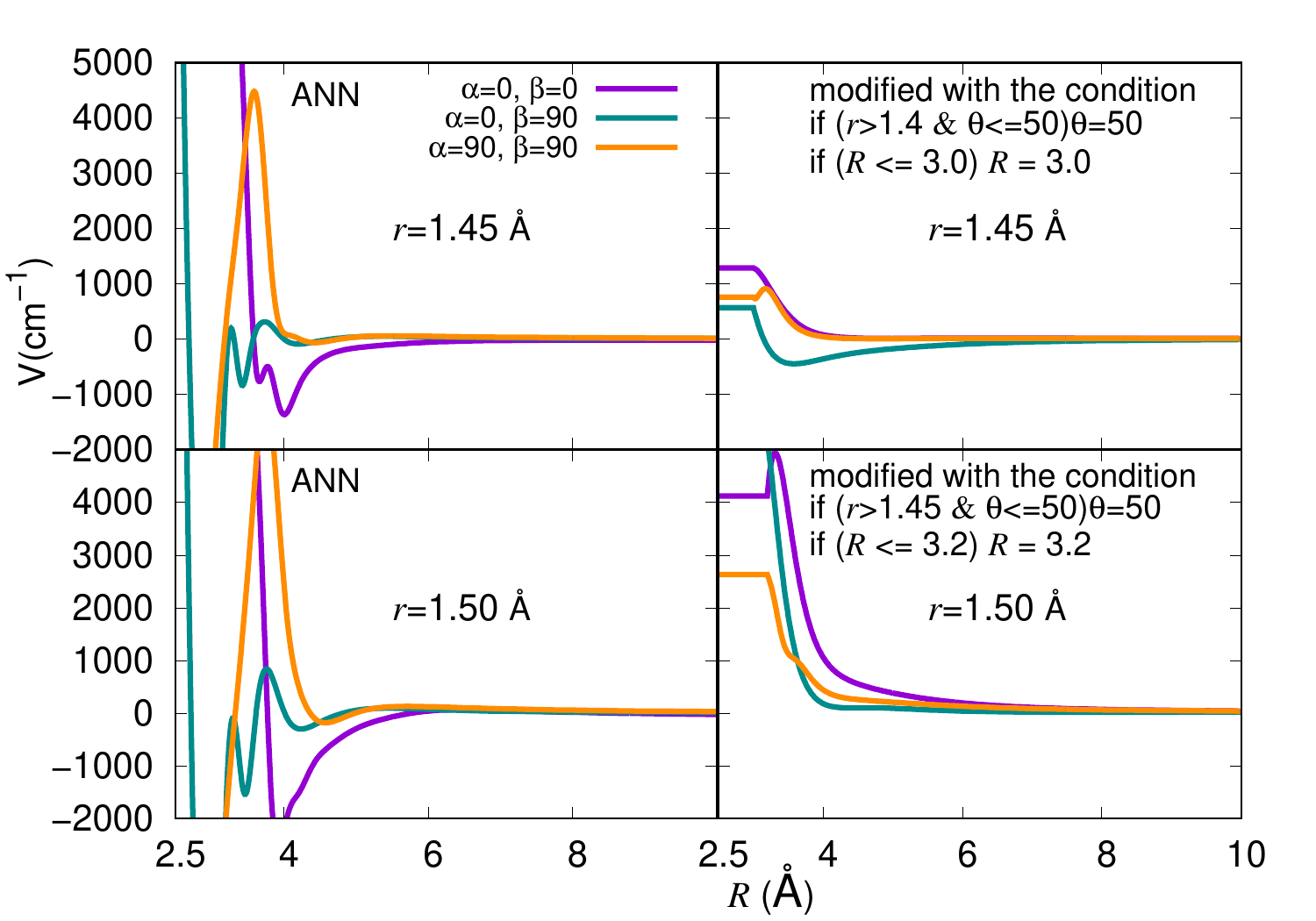} 

 \label{fig3c}
	\includegraphics[width=1.0\linewidth,angle=+0.0]{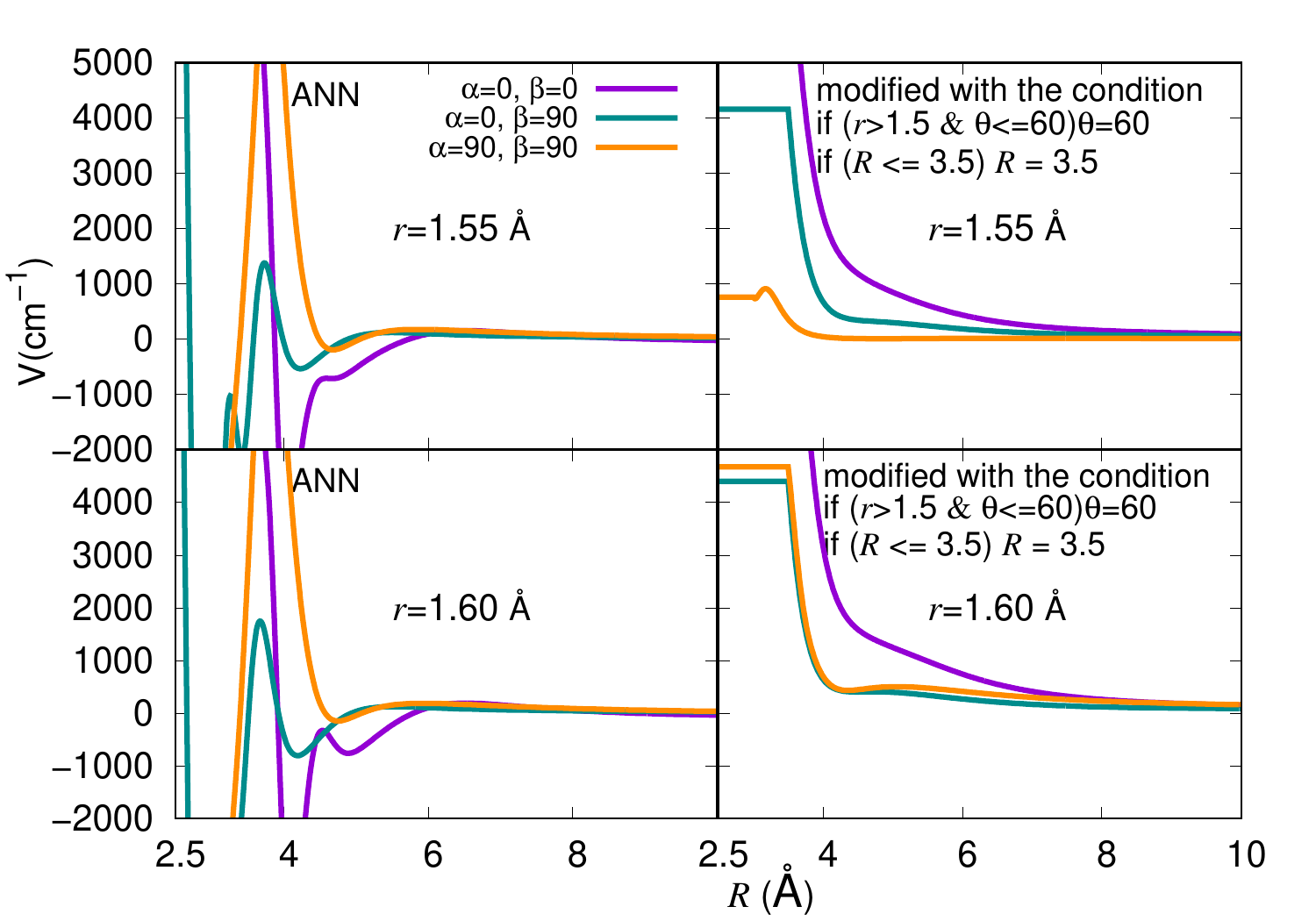} 
 
\caption{Graphical presentation of the results on the constrained ANN fitting for different bond lengths of the anion and for different selections of the three angles of the full 5D PES. See main text for further details.}
\label{fig3}

\end{figure}

 Since C$_2^-$ does not have a permanent dipole moment, the dominant terms in the long range potential are related to the polarizability of H$_2$ in its equilibrium bond distance: 
 
\begin{equation}
V_{\textnormal{LR}} = - \frac{\alpha_0}{2R^4} + \left(\frac{2\alpha_2}{R^4}\right)P_2(\textnormal{cos}\alpha), 
\label{eq:lr}
\end{equation}
where $\alpha_0$ = ($\alpha_{\parallel}$ + 2$\alpha_{\perp}$)/3 and $\alpha_2$ = $\alpha_{\parallel}$ - $\alpha_{\perp}$ with  $\alpha_{\parallel}$ and $\alpha_{\perp}$ being the parallel and perpendicular components, respectively, of H$_2$. The angle $\alpha$ is the  orientation angle given in Figure 1.

The polarizability for H$_2$ at $r_2$ = 1.4 $a_{0}$ (0.7408 \AA)  is given by $\alpha_{0}$ = 5.1786 $a_{0}^3$ and $\alpha_2$ = 1.8028 $a_{0}^3$ \citep{23NWL}. For comparison, $\alpha_{0}$ = 1.41 $a_{0}^3$ for He  (see: (\cite{LGS2023})).

We can therefore employ the above expression of the  long-range extension of the interaction, the $V_{\textnormal{LR}}$, to bring the ANN fitted 5D PES to larger radial distances. We  show in Figure 4 the effects of such a propagation to larger distances for the present ANN potential fit. We see from the reported examples in the two panels that to merge the two sets of points at distances of the order of about 24.0 \AA\ provides a smooth transition to the interaction at long distances, as shown by the curves in yellow labelled VNNLR.
   
The final form of the potential ($V_f$) using a switching function is given by,

\begin{equation}
    V_f = f_s V_{\textrm{ANN}}+(1-f_s)V_{\textrm{LR}}
\label{eq:vf}
\end{equation}
where the switching function $f_s(R)$ is,

\begin{equation}
    f_s(R) = \frac{1}{e^{\frac{(R-R_0)}{\Delta x}+1}}
\label{eq:sf}
\end{equation}
and $R_0 = 20.5 $ \AA, and $\Delta x$ = 0.5  \AA.

\begin{figure}
\centering
\includegraphics[width=1.0\linewidth, angle=+0.0]{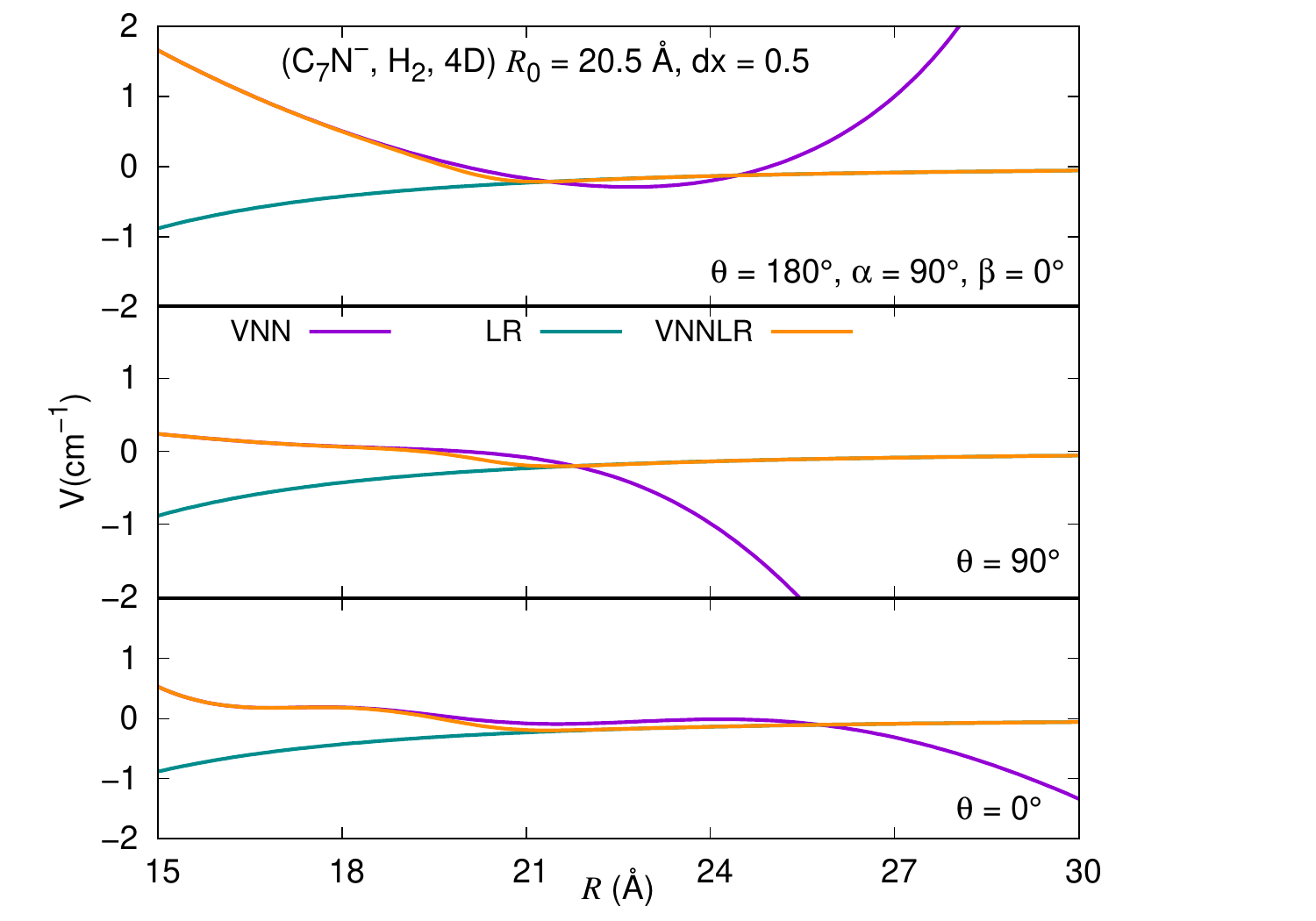}
        \caption{ Graphical presentation of the extended range of the 5D interaction using the LR formulae reported in eq.s (1), (2) and (3) in the main text. See there for further details.}
        \label{fig: Fig4}
\end{figure}

\subsection{Calculations of the coupling vibrational matrix elements within the 5D PES}
\label{sec:Vibr}

The scattering calculations described in the next section require  the interaction potential to be represented using an extended angular basis 
 as shown below:
 
\begin{equation}
V(R, r_1, \theta, \alpha, \beta) = \hspace*{-3mm} \sum_{\lambda_1,\lambda_2,\lambda} \hspace*{-3mm} A_{\lambda_1,\lambda_2,\lambda}(r_1, R) 
\times
Y_{\lambda_1,\lambda_2,\lambda}({\textbf{\^r}}_1, {\textbf{\^r}}_2, {\textbf{\^R}}),
\label{eq.vib_coup}
\end{equation}
\\
where $Y_{\lambda_1,\lambda_2,\lambda}
({\textbf{\^r}}_1, {\textbf{\^r}}_2, {\textbf{\^R}})$
 is a bi-spherical harmonics function given by:

\begin{eqnarray}
Y_{\lambda_1,\lambda_2,\lambda}({\textbf{\^r}}_1, {\textbf{\^r}}_2, {\textbf{\^R}}) = \hspace*{-5mm} \sum_{m_{\lambda_1},m_{\lambda_2},m_{\lambda}} \hspace*{-5mm}
\langle \lambda_1,m_{\lambda_1},\lambda_2,m_{\lambda_2}|\lambda,m_{\lambda}\rangle \nonumber \\
\times\ Y_{\lambda1}^{m_{\lambda_1}}({\textbf{\^r}}_1) Y_{\lambda_2}^{m_{\lambda_2}}({\textbf{\^r}}_2)
Y_{\lambda}^{m_{\lambda}}({\textbf{\^R}}),
\label{eq.harmonics}
\end{eqnarray}
%
where 0 $\leq \lambda_1 \leq$ 8, 0 $\leq \lambda_2 \leq$ 4, and only even values of $\lambda_1$ and $\lambda_2$ are
retained due to the symmetry of C$_2^-$ and H$_2$.

The next step in the representation of the interaction potential requires to construct matrix elements coming from the convolution of the potential in eq.(2) over the asymptotic vibrational wave functions of the C$_2^-$ partner

\begin{equation}
V_{\nu \nu'}(R,\theta,\alpha,\beta) =  \langle\chi_{\nu}(r_1)|V(R, r_1, \theta, \alpha, \beta)|\chi_{\nu'}(r_1)\rangle.
\label{eq.coupling}
\end{equation}
The off-diagonal $V_{01}(R,\theta,\alpha,\beta)$ term is the coupling which directly drives vibrationally inelastic $\nu = 1$ to $\nu=0$ transitions. At short ($R$) distances, the coupling terms are repulsive, becoming negligible quickly at longer distances, as is the case for many other atom-diatom systems. This can also be seen  for C$_2^-$ interacting with the H$_2$ partner as a point-like partner presented in our earlier work \cite{23NWL}. We shall further discuss this point below, where we show that the close-coupling scattering calculations discussed in the next section require the vibrationally-labelled coupling matrix elements to be further written in the form of a multi-pole expansion as

\begin{equation}
V_{\nu\nu'}(R,\theta,\alpha,\beta) = \sum_{\lambda_1,\lambda_2,\lambda}^{\lambda_{12\rm{max}}} V_{\nu\nu'}^{\lambda_1,\lambda_2,\lambda}(R) Y_{\lambda_1,\lambda_2,\lambda }({\textbf{\^r}}_1, {\textbf{\^r}}_2, {\textbf{\^R}})
\label{eq.harmonics}
\end{equation}
where again due to C$_2^-$ being a homonuclear diatomic, only even $\lambda$ terms are required. The bispherical-harmonic functions are defined in eq.(4) above.

To carry out additional tests on the reliability of the present evaluation of the vibrational coupling matrix elements, we have tried to reproduce the reduced-dimension results from our earlier work \cite{23NWL} where the 3D PES was employed, as discussed earlier in the Introduction Section, by using instead the calculated matrix elements with the full 5D interaction but scaling them within the 3D representation of the earlier work. A comparison of the present results with those from \cite{23NWL} is presented in the two panels of Figure 5. We show in its lower panel the off-diagonal expansion coefficients $V_{01}^{\lambda}$, for the first three values of the single $\lambda$ expansion index used in 3D: each of them compacts the three different indices for the 5D case that we have discussed before. The results in the upper panel compare the same type of results but for the diagonal matrix element $V_{00}^{\lambda}$.
We notice, first of all, that all the terms in the lower panel quickly approach zero as $R$ is
increased, indicating the essentially short-range nature of the vibrational coupling matrix elements between the two lower levels.
Additionally, all three multi-polar coefficients of the  $V_{01}^{\lambda}(R)$  matrix element are steeply repulsive as $R$ decreases. The solid lines report the 3D coefficients reproduced via the present ANN expansion, while the dashed lines are our earlier results from the 3D calculations. On the whole, we can see that the general shape of these coupling matrix elements follows what we had found in the earlier study and are  well reproduced by the constrained representation of the 5D interaction once reduced to the 3D version.

The diagonal matrix elements given by the upper panel show  some quantitative  differences between the present findings (solid lines) and the earlier results (dashed lines): the scaled 5D diagonal terms are somewhat smaller in strength than the earlier calculations in 3D. This result indicates that the use of a larger dimensional PES, as in the present case, distributes the anisotropic features over a larger angular range, which is therefore affected by the scaling into the reduced dimensionality. On the other hand, the off-diagonal coupling terms in the lower panel of Figure 5 act on a much shorter range of radial values and therefore sample a very similar angular anisotropy in both surfaces. This is a reassuring finding since our present study is chiefly directed to testing state-changing vibrational processes, dominated by the off-diagonal matrix elements of the coupling potential.

\begin{figure}
\centering

\label{fig5a}
	\includegraphics[width=0.70\linewidth,angle=-90.0]{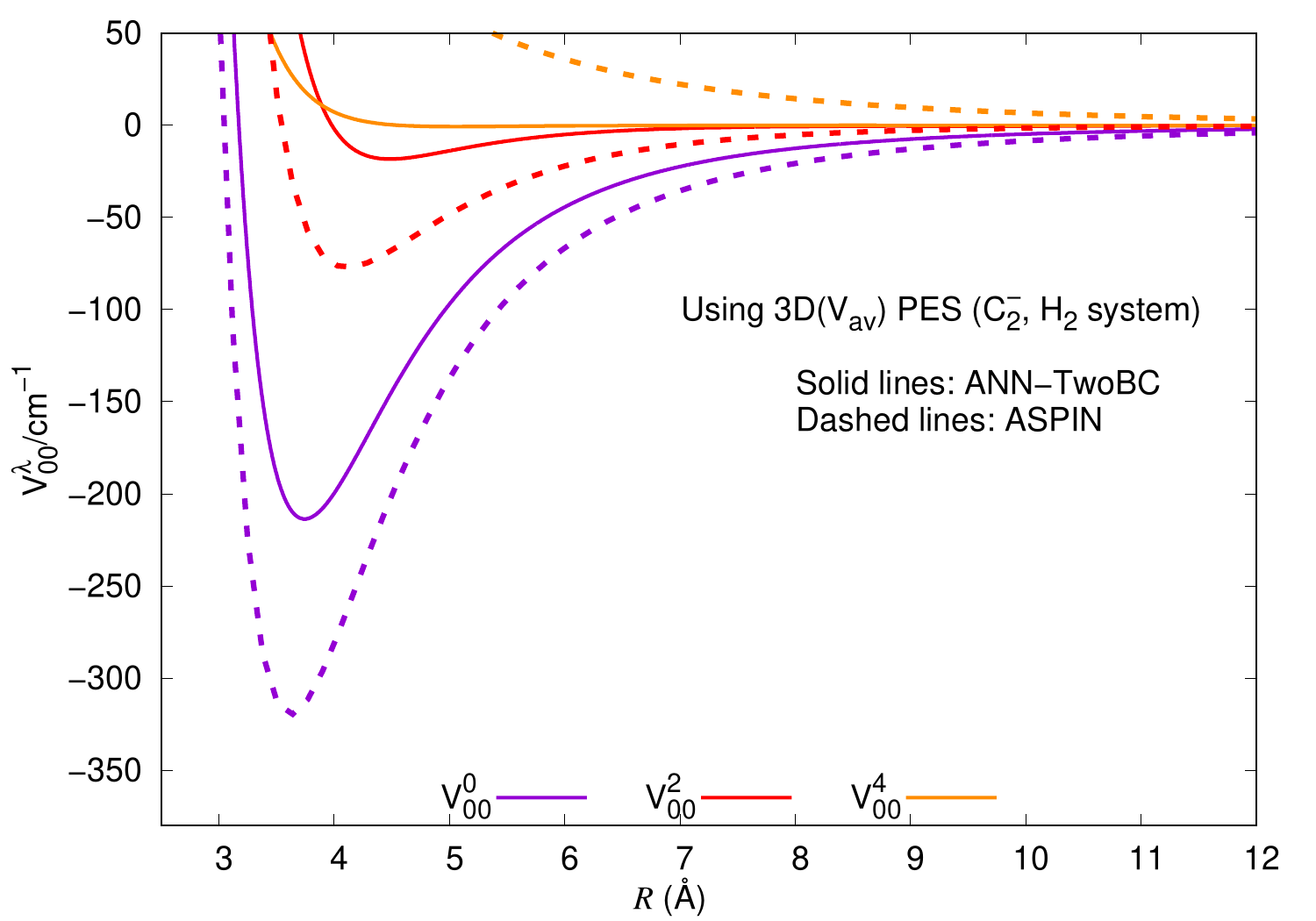}

\label{fig5b}
	\includegraphics[width=0.70\linewidth,angle=-90.0]{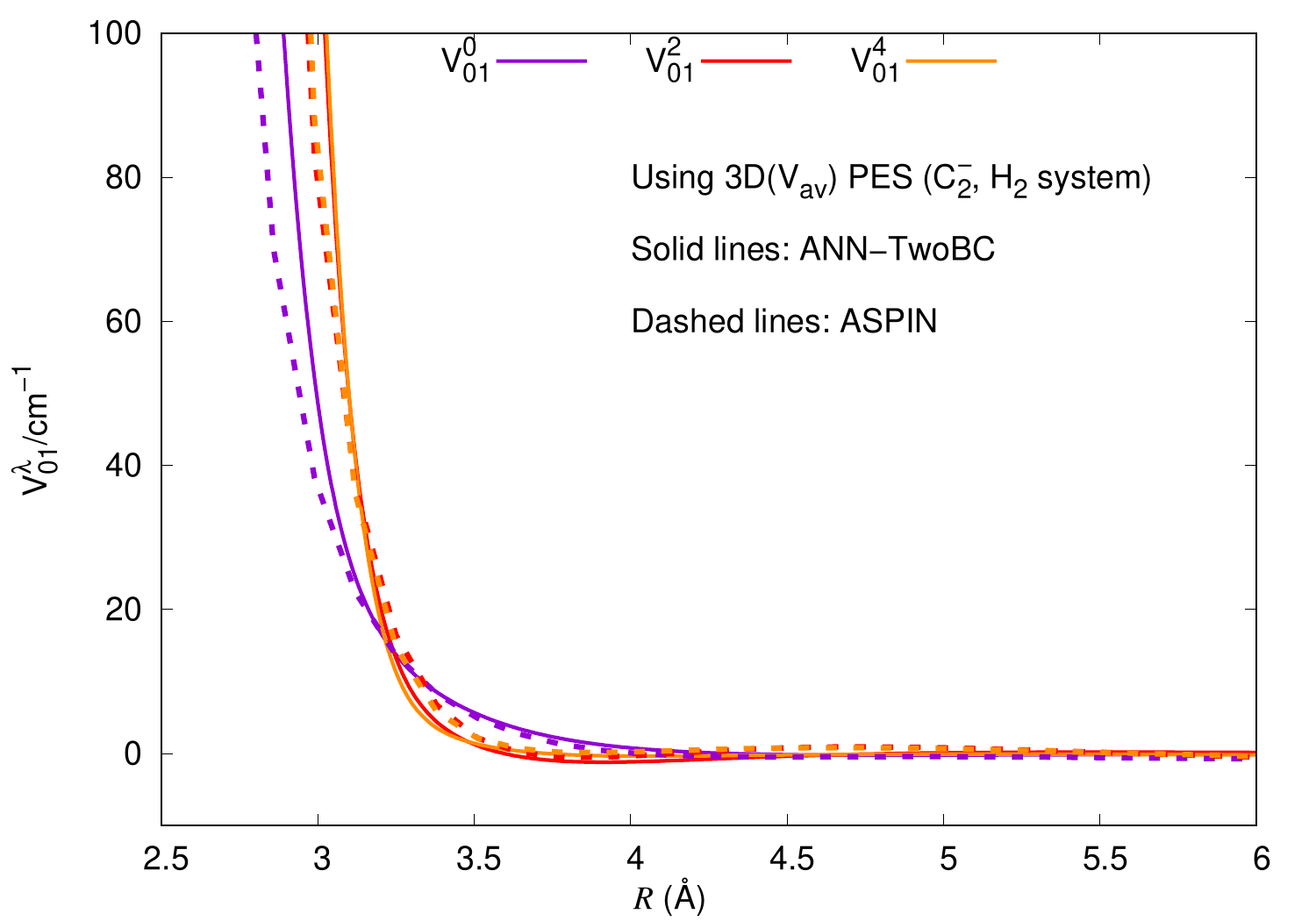}

\caption{Comparison of coupling matrix elements obtained from the present ANN fitting from the 5D original points (solid lines, using the TwoBC code) with those obtained from our earlier work of \cite{23NWL} (dashed lines, using the ASPIN code). Each curve corresponds to a different value of the expansion index $\lambda$, as a superscript and as given in eq.(5). Top panel: diagonal coupling between the $\nu$ = 0 levels. Lower panel:  coupling between the $\nu$ = 0 and $\nu$ = 1 levels. See main text for further details.}
\label{fig5}
\end{figure}

\section{Quantum scattering calculations} 
\label{sec:scat}

Quantum scattering calculations were carried out using the coupled channel (CC) method to solve the Schr\"{o}dinger equation for
scattering between two vibrating-rotating diatomic molecules as implemented in the code developed by R.V.Krems \cite{16RVK}.
The method has been described in detail before \cite{60ArDaxx,08LoBoGi} 
and only a brief summary will be given here, with the equations given in atomic units. 
For a given total angular momentum $\bf{J} = \bf{l} + \bf{j}_{12}$, where $\bf{j}_{12} = \bf{j}_1 + \bf{j}_2$, the latter two rotational angular momenta pertaining to the two interacting molecular rotors, while      $\bf{l}$ describes the relative angular momentum between partners.

Then, the scattering wave function is
expanded as

\begin{eqnarray}
\Psi^{JM}(R, r_1, \theta, \alpha, \beta) = \frac{1}{R}\sum_{\nu,k}f^{J}_{\nu k}(R)\chi_{\nu,j_1}(r_1) \nonumber\\
\times Y^{JM}_{k}({\textbf{\^r}}_1, {\textbf{\^r}}_2, {\textbf{\^R}})     
\end{eqnarray}
where $k = \{j_1,j_2,j_{12},l\}$.

 The $\chi_{\nu,j_1}(r_1)$ are the 
radial part of the ro-vibrational eigenfunctions of the anionic molecule taking part in the collisions, while
the $f^{J}_{\nu k}(R)$ are
the radial expansion functions which need to be determined: see reference \cite{16RVK}.

Substituting the expansion into the Schr\"{o}dinger equation 
with the Hamiltonian for diatom-diatom scattering, \cite{16RVK, 60ArDaxx,08LoBoGi}, leads to the  CC equations for each $J$:

\begin{equation}
\left(\frac{d^2}{dR^2} + \mathbf{K}^2 - \mathbf{V} - \frac{\mathbf{l}^2}{R^2} \right) \mathbf{f}^J = 0.
\label{eq.CC}
\end{equation}
Here each element of $\mathbf{K} = \delta_{i,j}2 \mu (E- \epsilon_i)$ (where $\epsilon_i$ is the channel asymptotic energy),
$\mu$ is the reduced mass of the system, $\mathbf{V}= 2 \mu \mathbf{U}$ is the interaction potential matrix between
channels and $\mathbf{l}^2$ is the matrix of orbital angular momentum. For the ro-vibrational scattering calculations of interest here,
the matrix elements $\mathbf{U}$ are given explicitly as

\begin{eqnarray}
\langle \nu j k J | V | \nu' j' k' J \rangle = \int_0^{\infty} \mathrm{d}r_1 \int \mathrm{d} \hat{\mathbf{r_1}}\int \mathrm{d} \hat{\mathbf{r_2}}\int \mathrm{d} \hat{\mathbf{R}} 
\nonumber \\
\chi_{\nu,j_1} (r_1)Y^{JM}_{k}
({\textbf{\^r}}_1, {\textbf{\^r}}_2, {\textbf{\^R}})V(R,r_1,\theta,\alpha,\beta) \chi_{\nu',j_1'}^*(r_1)  
\nonumber \\
\times Y^{JM}_{k}
({\textbf{\^r}}_1, {\textbf{\^r}}_2, {\textbf{\^R}}),
\label{eq.vib_elements}
\end{eqnarray}
%
where the angular functions are those defined right after eq.(4).

Since the intermolecular potential $V(R,r_1,\theta,\alpha,\beta)$ is expressed as in  Eq.(7), the matrix elements  can now also be written as

\begin{equation}
\langle \nu j_1 k J | V | \nu' j_1' k' J \rangle = \sum_{\lambda_1,\lambda_2,\lambda}^{\infty} V_{\nu \nu'}^{\lambda_1,\lambda_2,\lambda}(R) f^J_{\lambda_{12} j_1 k j_1' k'},
\label{eq.vib_elements2}
\end{equation}
\\
where the $f^J_{\lambda_{12} j_1 k j_1' k'}$ terms are the Percival-Seaton coefficients given as

\begin{eqnarray}
f^J_{\lambda_{12} j_1 k j_1' k'} = \int\mathrm{d} \hat{\mathbf{r_1}}\mathrm{d}\hat{\mathbf{r_2}}\int \mathrm{d} \hat{\mathbf{R}} \times Y^{JM}_{k}
({\textbf{\^r}}_1, {\textbf{\^r}}_2, {\textbf{\^R}})^* \nonumber\\
P_{\lambda}(\cos \theta) \times Y^{JM}_{k}
({\textbf{\^r}}_1, {\textbf{\^r}}_2, {\textbf{\^R}})),
\label{eq.Percival-Seaton}
\end{eqnarray}
\\ \\
for which analytical forms are known \cite{08LoBoGi}. Eq. \ref{eq.vib_elements2} also makes use of the widely known approximation
\begin{equation}
V_{\nu,\nu'}^{\lambda_1,\lambda_2,\lambda}(R) \approx V_{\nu j_1 \nu' j_1'}^{\lambda_1,\lambda_2,\lambda}(R),
\label{eq.vib_approx}
\end{equation}
for all $j_1$ such that the effect of rotation on the vibrational matrix elements is ignored. Their actual excitation within the vibrational cooling dynamics will however be treated exactly.

The CC equations are propagated outwards from the
classically forbidden region to a sufficiently large distance where the scattering matrix $\mathbf{S}$ can be obtained. The actual numerical details will be further given below. 

The 
ro-vibrational state-changing cross sections are obtained as
\begin{eqnarray}
\sigma_{\nu j_1j_2 \rightarrow \nu^{'}j_1^{'}j_2^{'}} (E_{\textnormal{coll}}) = \frac{\pi}{(2j_1 + 1)(2j_2 + 1)k^2_{vj_1j_2}}\sum_{J}(2J + 1) \nonumber\\ 
\times 
\sum_{j_{12},j_{12}^{'},l,l^{'}}|\delta_{\nu j_1j_2lj_{12},\nu^{'}j_1^{'}j_2^{'}l^{'}j_{12}^{'}}- \hspace*{5mm}\nonumber\\
S^J_{\nu j_1j_2lj_{12},\nu^{'}j_1^{'}j_2^{'}l^{'}j_{12}^{'}}(E_{\textnormal{coll}})|^2, 
\hspace*{10mm}
\end{eqnarray}
where $E_{\textnormal{coll}}$  is the collision energy.

In all scattering calculations the C$_2^-$ anion was treated as a pseudo-singlet ($^1 \Sigma$) and the effects of 
spin-rotation coupling were ignored. In our previous work on this system it was shown that a pseudo-singlet treatment of the rotational 
state-changing collisions resulted in essentially the same results as the explicit doublet calculation when the relevant cross sections 
were summed \cite{20MaGiGo}. This approximation reduces the computational cost of the scattering calculations without significantly 
affecting the size of the cross sections and thus the present conclusions.

For the CC equations to converge, a rotational basis set was used which included up to $j_1$ = 20 rotational functions for each vibrational state and for either the ortho-H$_2$ or para-H$_2$ collision partners. 
The CC equations were propagated along $R$ between 1.7 and 1500.0 {\AA} using the Johnson's log-derivative matrix propagation \cite{86Maxxxx.c2m} . The potential energy was  calculated to generate the 
$V_{\nu\nu'}^{\lambda_1,\lambda_2,\lambda}(R)$ values using our long-range correction discussed earlier and then employed directly as such within the TwoBC code employed in our calculations \cite{16RVK}. Our \textit{ab initio} calculated interaction energies were computed to $R = 41$ \AA $ $, a distance where the
interaction energy is expected to be fairly negligible for the temperature of interest here. In any case, we carried out the potential extrapolation using our long-range terms to much larger $R$ values (see above) to make sure that by that distance the extrapolated PES would have a  negligible effect on the computed
cross sections \cite{20MaGiGo}.

A number of parameters of the calculation were checked for convergence. The number of $\lambda_1,\lambda_2,\lambda$ terms from Eq.(10)
was checked for both rotationally and vibrationally inelastic collisions. For the former, calculations were converged to better than 
1 \% using only three terms (up to $\lambda = 4)$. For vibrationally inelastic collisions the convergence with increasing $\lambda_1,\lambda_2,\lambda$ are less
precise.  This is due to the very small cross sections for these processes which makes obtaining precise and stable values 
more difficult to achieve. For production calculations, more than 20 ($\lambda_1,\lambda_2,\lambda$) terms were included for each $V_{\nu\nu'}(R)$.

Despite this practical difficulty, the vibrationally inelastic cross sections remained reasonably consistent and thus our $R$ range is sufficiently large  
to obtain cross sections which are to the correct order of magnitude, sufficient to assess rates for vibrational 
quenching of C$_2^-$ in collision with H$_2$.

As a final check of our calculation parameters, the effect of the vibrational basis set was also considered. In all calculations we used the
vibrational energies and rotational constants obtained from calculations using LEVEL, as discussed in \cite{23NWL}, and employing our own C$_2^-$ potential energy curve as discussed in our earlier work \cite{23NWL}. It was found that for $\nu = 0$ and $\nu = 1$, which are the  states of interest here (see next section),
it was sufficient to only include these two asymptotic vibrational states.

Scattering calculations were carried out for collision energies between 1 and 1500 cm$^{-1}$ using steps of 0.1 cm$^{-1}$ for 
energies up to 11 cm$^{-1}$, 0.2 cm$^{-1}$ for 100-200 cm$^{-1}$, 1.0 cm$^{-1}$ for 11 to 50 cm$^{-1}$, 2 cm$^{-1}$ for
50 to 100 cm$^{-1}$ and 5 cm$^{-1}$ for 100 to 200 cm$^{-1}$.   Then we used steps of 10.0 cm$^{-1}$for intervals from 200 to 400 cm$^{-1}$, then steps of  20.0 cm$^{-1}$for  the range from 400 to 700 cm$^{-1}$, steps of 50 cm$^{-1}$ for the range from 700 to 1000 cm$^{-1}$ and steps of  100 cm$^{-1}$ for the final energy range from 1000 to 1500 cm$^{-1}$.
This fine energy grid was used to ensure that important features such as 
resonances appearing in the cross sections were accurately accounted for and their contributions correctly included when the corresponding
rates were calculated. The number of partial waves was increased with increasing energy up to $J = 80$ for the highest energies considered.

\section{Results and discussion}
\label{sec:results}

\subsection{Ro-vibration inelastic cross sections of the anionic partner}
\label{sec:rot_xsec}

To analyse as extensively as possible the differences in the dynamical behaviour of the title system when the full dimensionality of the 5D PES is employed, we start with presenting in Figure 6 the computed inelastic cross sections using both the 5D interaction potential and the 3D interaction discussed in our earlier work \cite{23NWL}. The different colouring for the two sets of calculations is indicated in that Figure's caption.

The two sets of four panels of Figure 6 report eight different rotational  transitions accompanying the vibrational de-excitation for the anionic rotor, while the para-H$_2$ partner is left in its ground rotational state ($j_2'$ = 0). The main vibrational state-changing process is the same for all the reported curves and therefore not indicated in their labels: it involves the $\nu = 1$ to $\nu = 0$ de-excitation process. Firstly, we clearly see in all panels that using the more accurate 5D representation of the interaction produces inelastic cross sections that are invariably larger than those obtained in our previous work using the 3D interaction. This is particularly evident at higher collision energies, where the size differences can be as large as more than two orders of magnitude. Additionally, we see in the low-energy range that a large number of narrow resonant structures appear, as expected for the present interaction between light partners. Since our primary interest in the present work involves the comparative sizes of the inelastic rate coefficients, discussed later in the paper, we have not further carried out a detailed analysis of such resonant features: none of them has been directly revealed in the existing experiments so far. On the other hand, we have made sure that our  energy grid was fine enough to include all the peak regions of the cross sections, since it is important to have them correctly represented for later calculating the inelastic rate coefficients (see below). It is significant to note that the top-left panel of Figure 6 indicates the purely vibrational de-excitation channel, since both colliding rotors are left in their ground rotational state. We clearly see that the purely vibrationally inelastic data are very close in size for both calculations, with the 5D results becoming larger than the 3D data at collision energies of about 10  cm$^{-1}$ and above.

\begin{figure}
\centering
\label{fig8a}
	\includegraphics[width=0.93\linewidth,angle=+0.0]{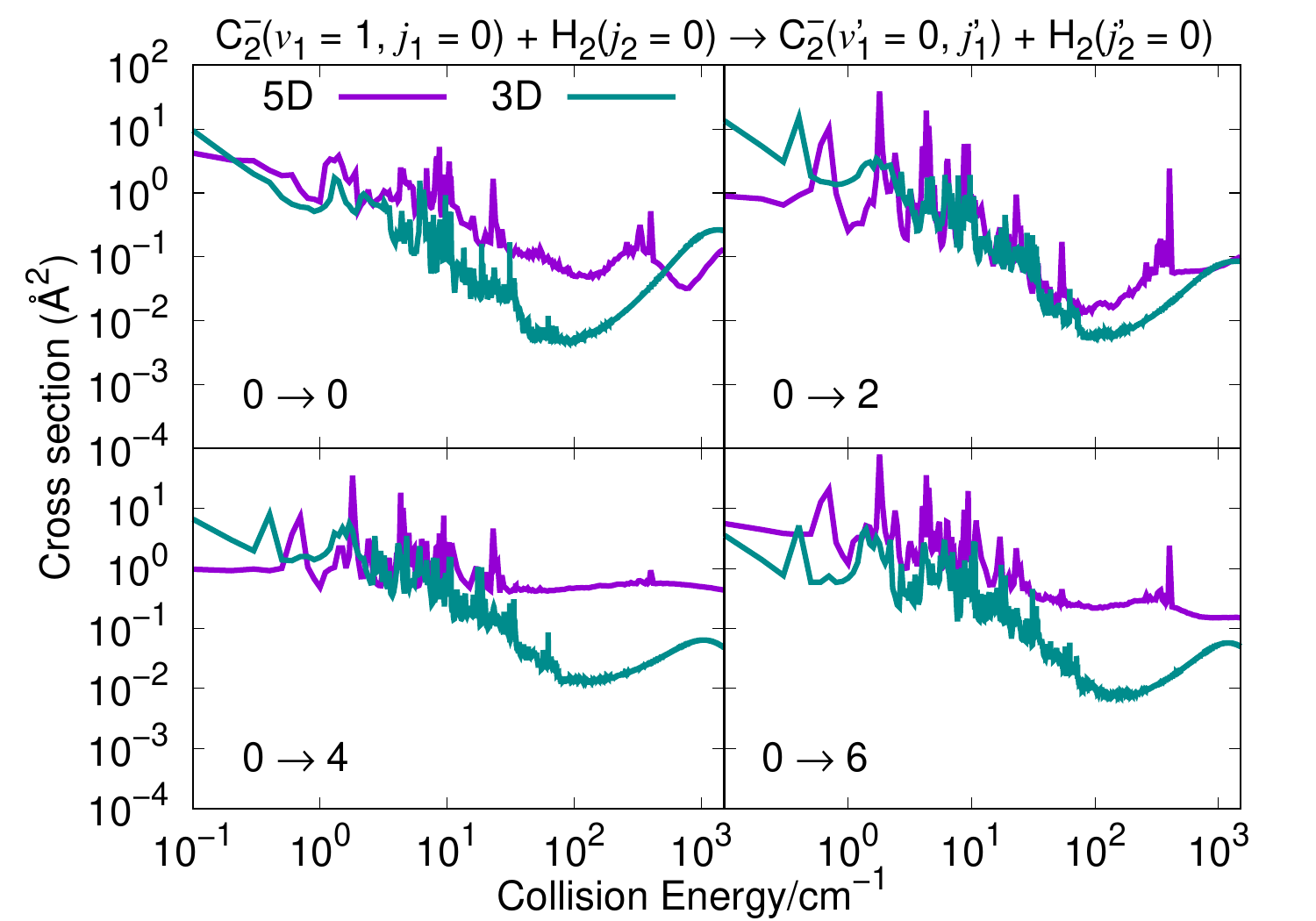}
\label{fig8b}
	\includegraphics[width=0.93\linewidth,angle=+0.0]{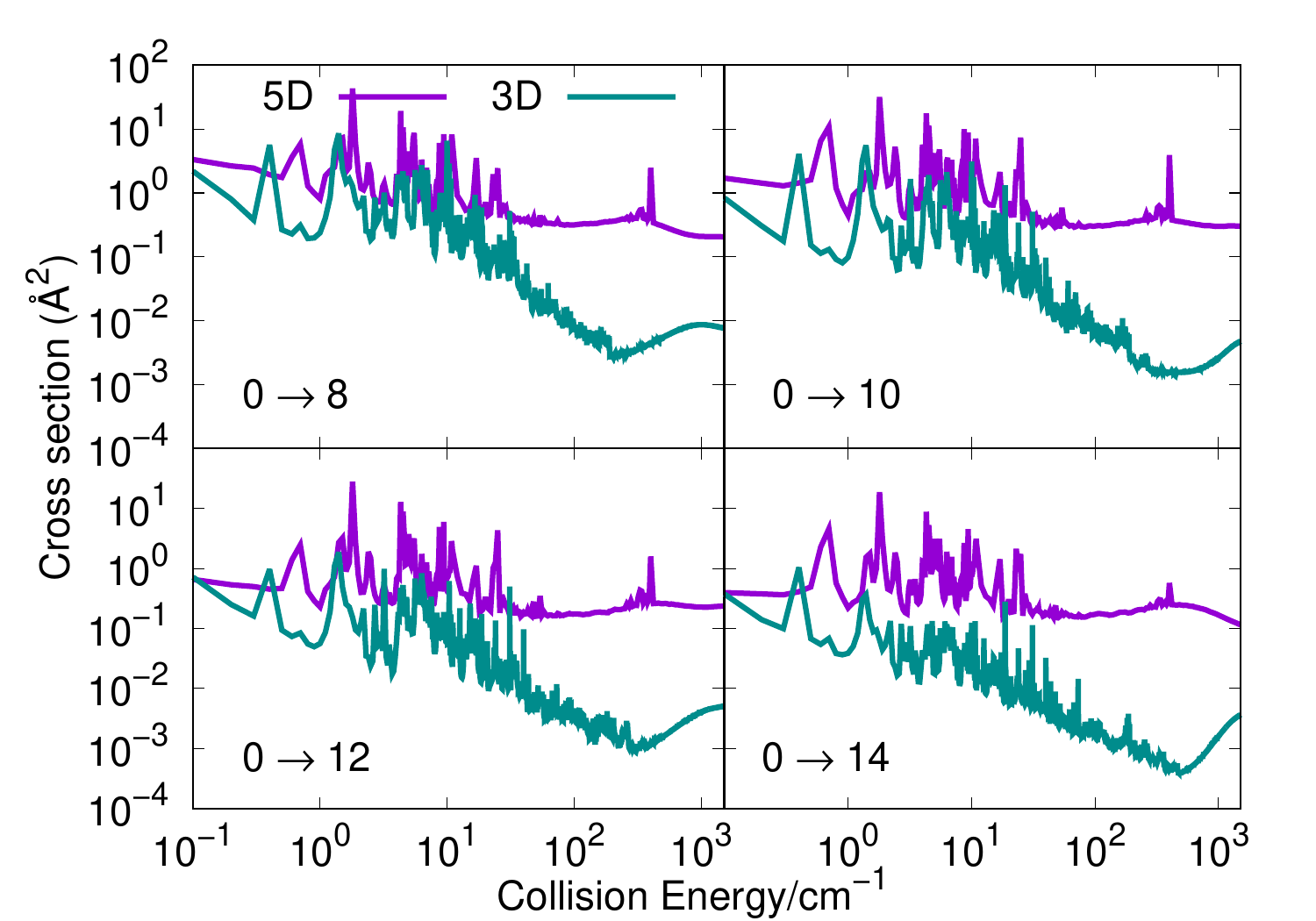}

\caption{Comparison of computed rotationally inelastic cross sections accompanying the vibrational de-excitation of the anion  using either the full 5D interaction potential (purple curves) or the reduced 3D interaction potential described in \cite{23NWL} (green curves).The collision partner is the para-H$_2$ neutral molecule. The labeling within each panel  refers only to the initial and final rotational states of the molecular anion before and after its vibrational cooling. The upper set involves transitions between the lowest four levels, while the lower set shows transitions to  four higher levels. See the main text for further details.}
\label{fig8}
\end{figure}

The lower four panels  of Figure 6 present the same set of processes as those discussed before, but involving this time the higher lying rotational states of the C$_2^-$ partner. One marked difference from the earlier data in Figure 6 is that although the 5D dynamics still produces larger cross sections (purple curves), they are more consistently so, especially in the low-energy range where differences of more than one order of magnitude can be observed. Figure 6 also makes clear the fairly small values of the involved inelastic processes, dominated by the vibrational relaxation channel, some of which become even smaller as the collision energy increases above 10 cm$^{-1}$. The present new data  show a broadening of the size range as one moves to higher collision energies, where we see that the larger the rotational energy transfer the smaller the corresponding cross sections.

A more directly descriptive view of the relative sizes of the various computed cross sections of Figure 6, as the final rotational state changes, can be seen in the sort of 'stick spectrum' reported by Figure 7. We present there three sets of panels, each of them showing the final state distribution at three different collision energies. Hence, the upper set of panels reports the $j_1$ = 0 as the initial state, while the middle set starts all transitions from $j_1$ = 2. Finally, the lower set of panels indicates $j_1$ = 4 as the initial state of all the transitions shown. The following comments can be made from a perusal of these results:

(i) the cross sections in the upper panels are all uniformly larger than those where the initial state of the anion becomes higher: $j_1$ = 2 and 4 in the middle and lower panels, respectively;

(ii) the reduction in size is most marked for the data in the lower set of panels, where the cross sections become smaller by nearly one order of magnitude on an average;

(iii) in nearly all the data shown in all three sets of panels we see that the largest excitation probabilities occur for different final rotational states depending on the initial state of the anion before the collision. On the other hand, the  amount of angular momenta transferred during the largest cross sections remains in general the same, peaking at  $\Delta$$j_1$ = 4 or $\Delta$$j_1$ = 6. The same occurs in all three sets of panels;

(iv) In general, we can also say that the transitions where more units of angular momenta are transferred to rotations yield increasingly smaller cross sections, which become nearly negligible by the time $j_1'$ becomes equal to 20.

\begin{figure}
\centering

\label{fig7a}
	\includegraphics[width=0.97\linewidth,angle=+0.0]{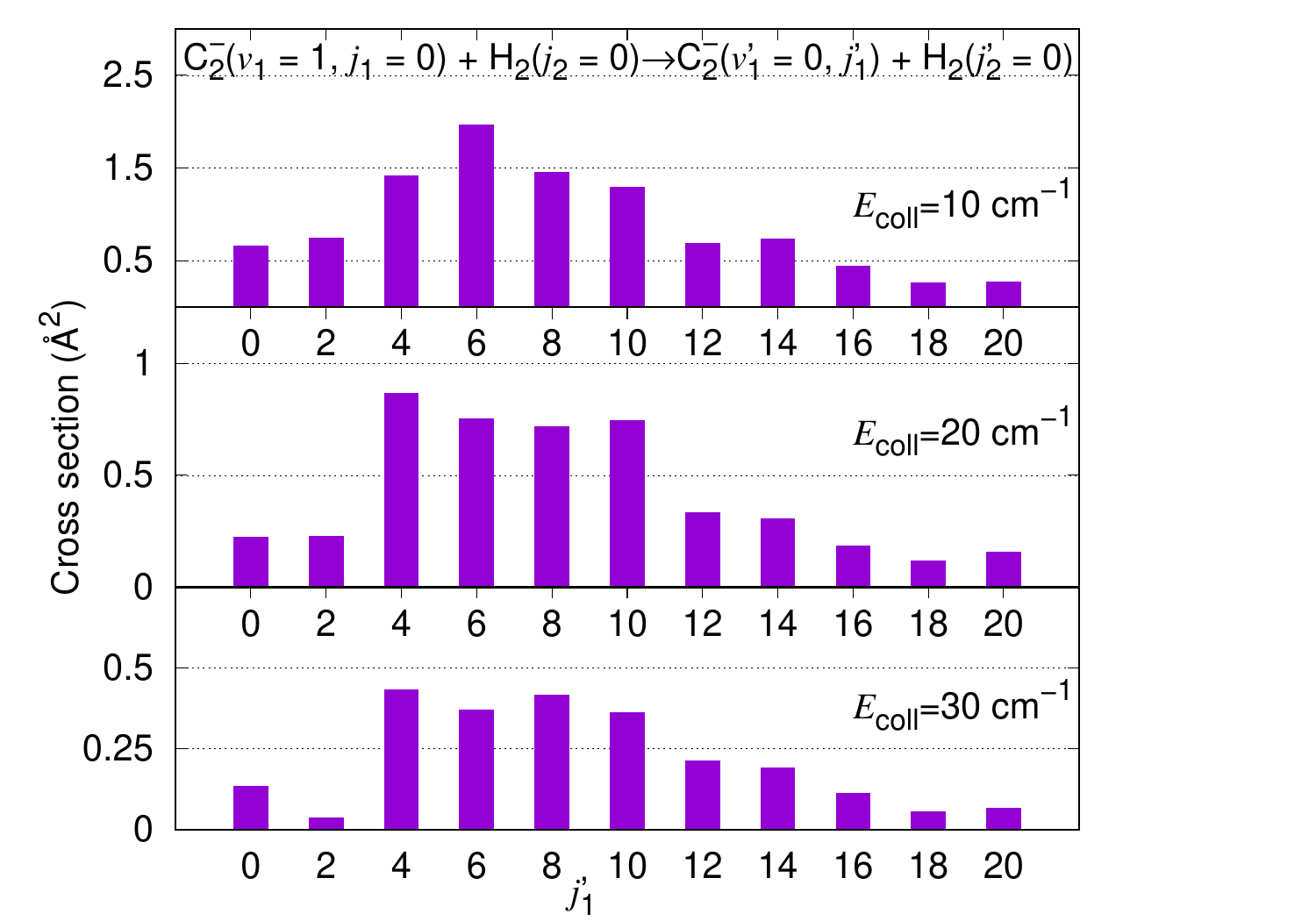}

\label{fig7b}
	\includegraphics[width=0.97\linewidth,angle=+0.0]{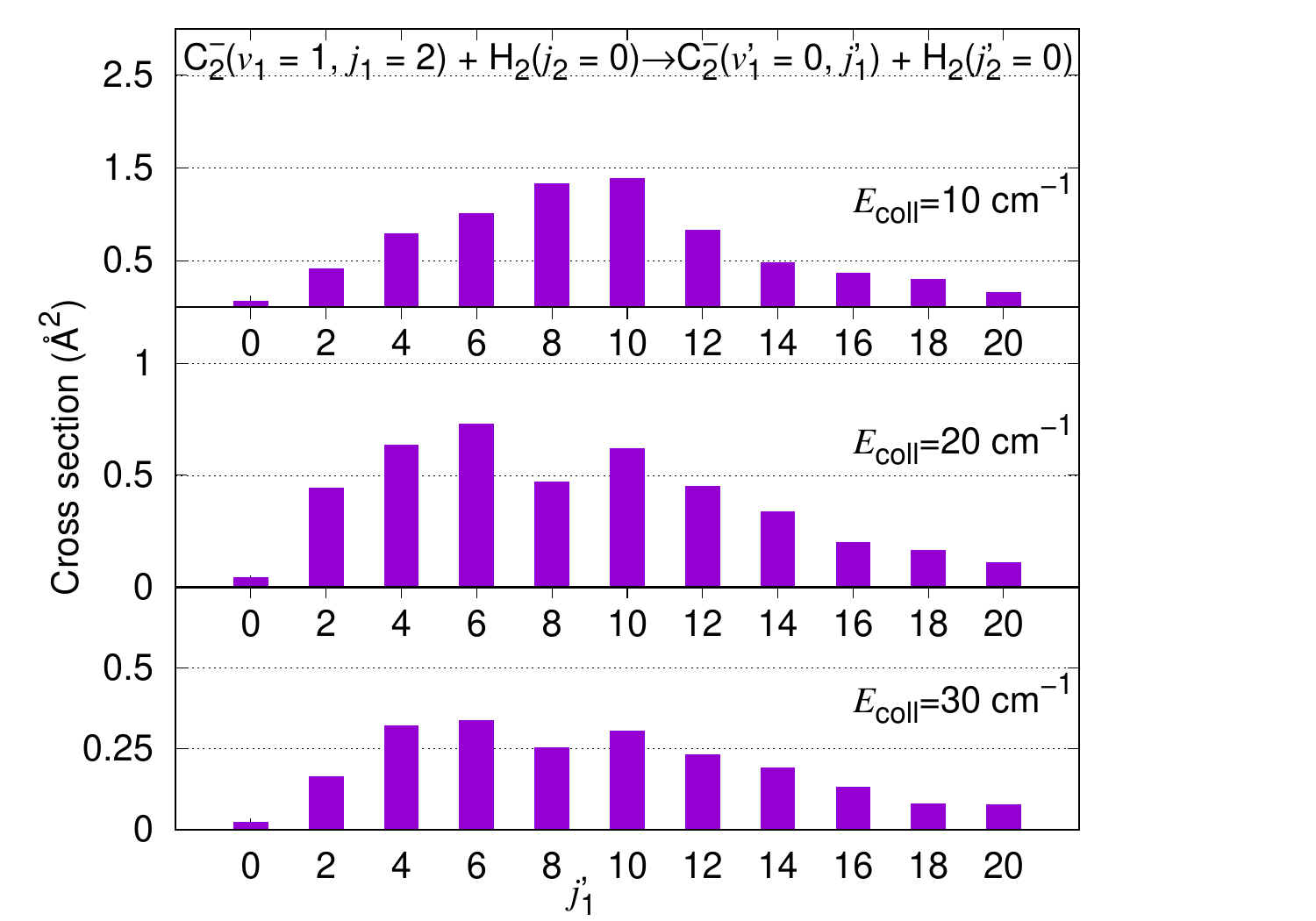}

\label{fig7c}
	\includegraphics[width=0.97\linewidth,angle=+0.0]{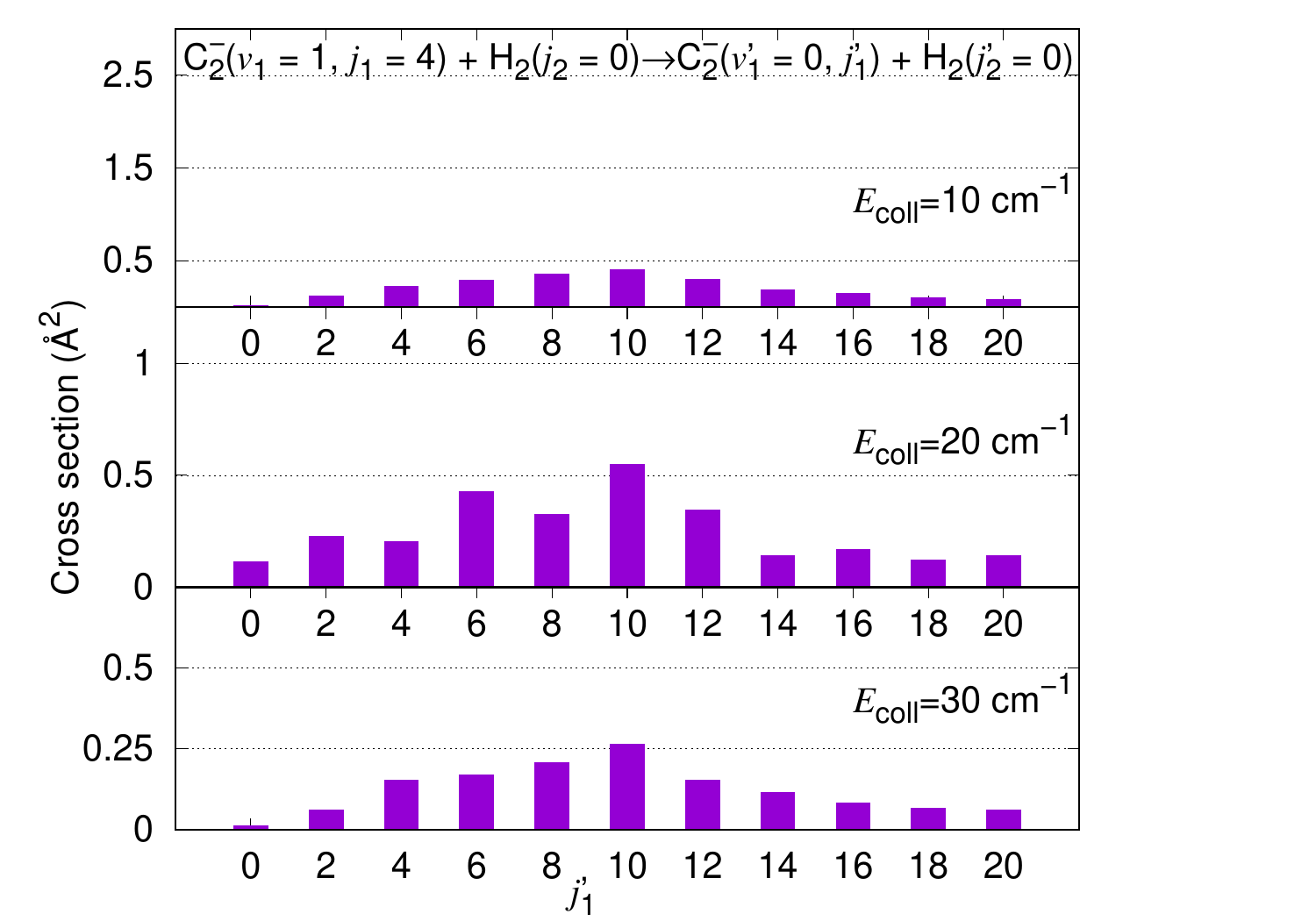}

 \caption{Computed inelastic cross sections as those given in Figure 6, but here presented as vertical bars for each final rotational state. The neutral molecular partner is again the para-H$_2$.  The initial state changes in each set of panels: (i) the upper panel has $j_1$ = 0 as the initial state, (ii) the middle panel has $j_1$ = 2; (iii) the lower panel has $j_1$ = 4. See the main text for further details.}
 
\label{fig7}
\end{figure}
\FloatBarrier

An interesting comparison could be obtained by repeating the calculations shown in the panels of Figure 6, but this time using the ortho-H$_2$ as the collision partner. In this case the ground rotational state of the neutral molecule is given by the $j_2$ = 1, hence we are dealing with a rotating partner that modifies the angular momenta coupling with respect to the case of the para-H$_2$, which is a non-rotating collision partner. The results given by Figure 8 compare our present calculations for the two cases.
\begin{figure}
\centering
\label{fig8a}
	\includegraphics[width=0.93\linewidth,angle=+0.0]{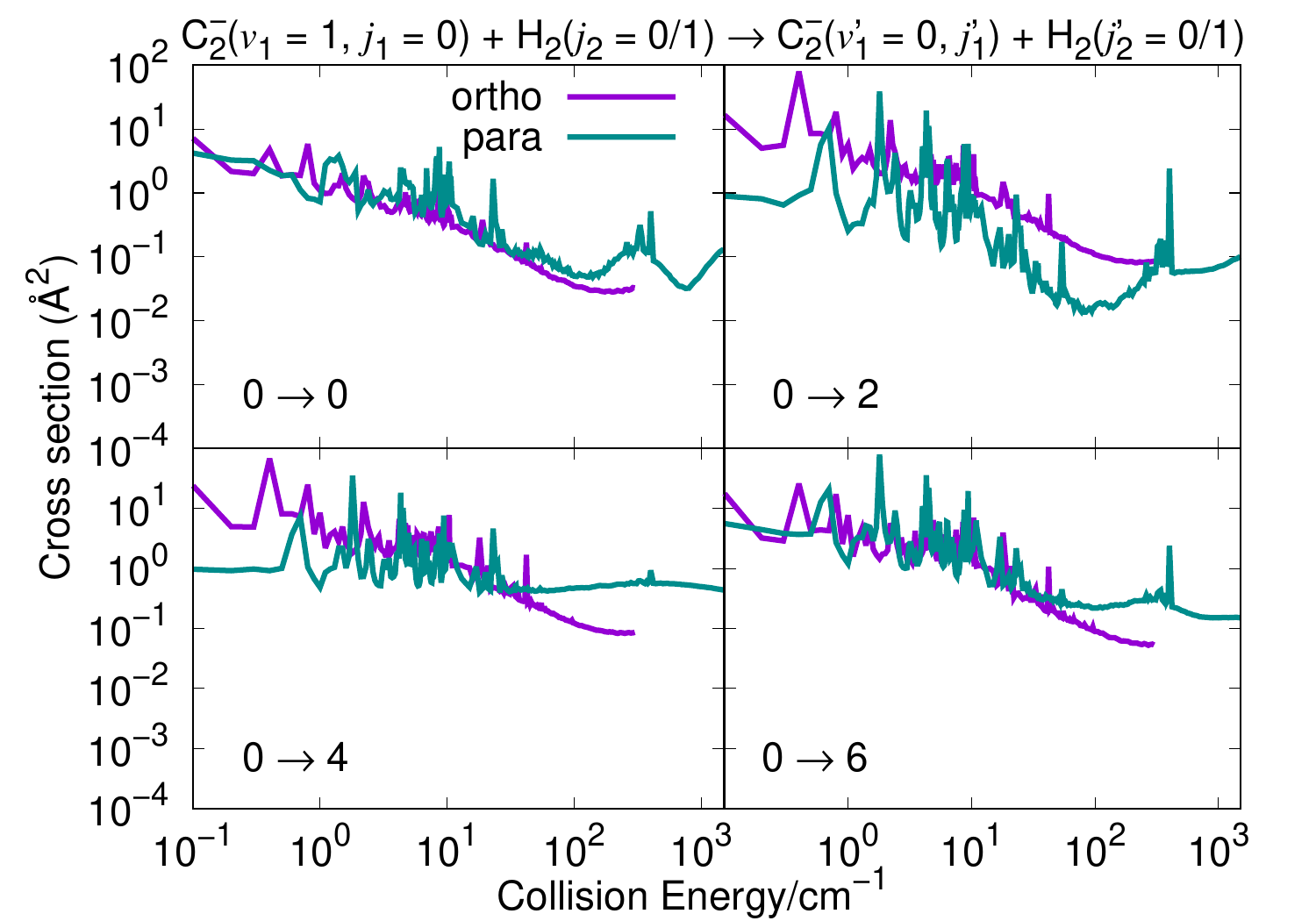}
\label{fig8b}
	\includegraphics[width=0.93\linewidth,angle=+0.0]{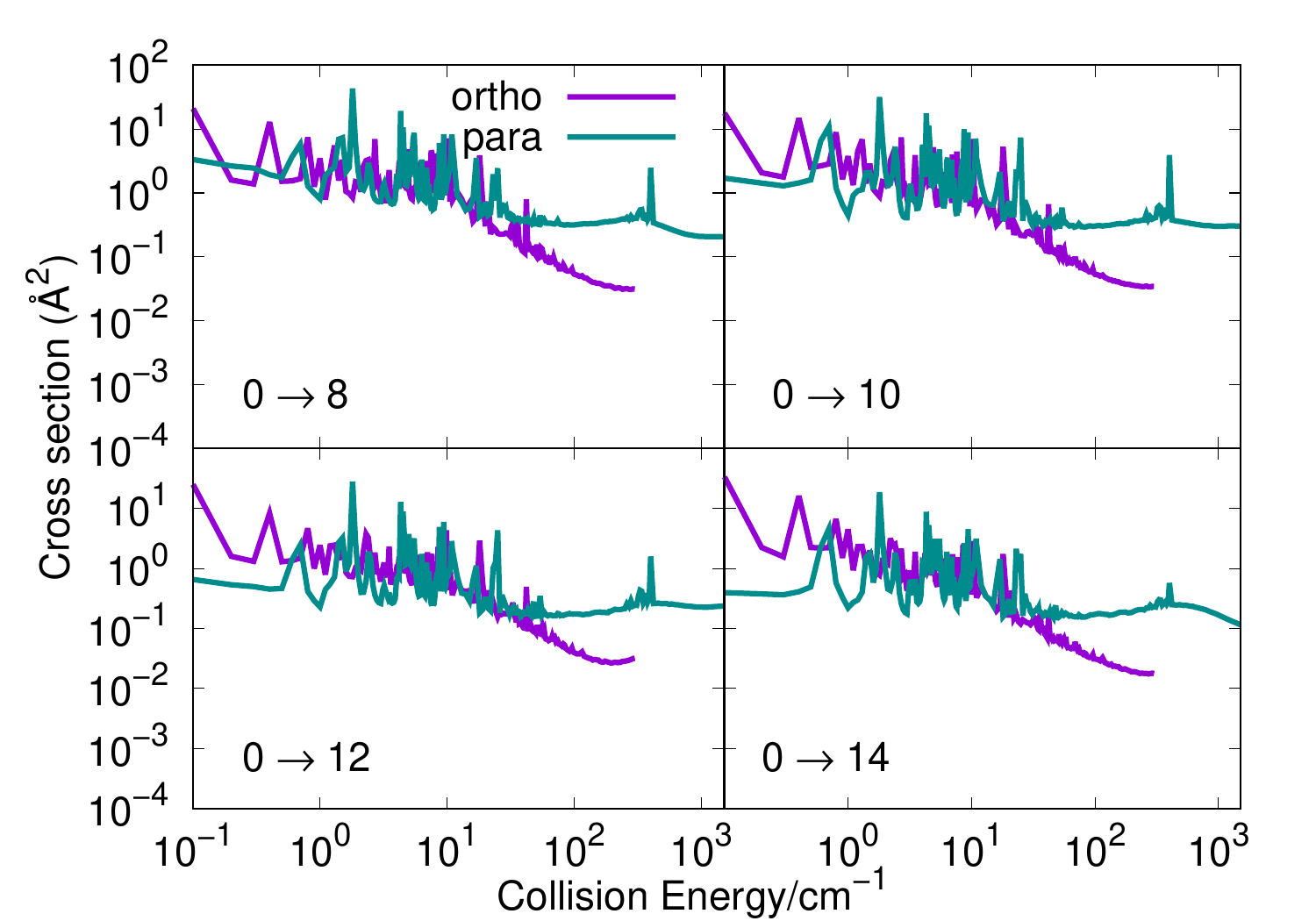}

\caption{Comparison of computed rotationally inelastic cross sections accompanying the vibrational de-excitation of the anion  using  the full 5D interaction potential (purple curves) with the ortho-H$_2$ as a partner or the earlier results using the para-H$_2$ as a collision partner. The labeling within each panel  refers only to the rotational states of the anion. The upper set involves transitions between the lowest four levels, while the lower set shows transitions to higher four levels. See the main text for further details.}
\label{fig8}
\end{figure}

The results reported in the two sets of panels show the strong similarities between the energy behaviour and the relative size of the two different calculations. In all cases we see that the ortho-cross sections are either very similar or slightly larger than the para-cross sections, this being the case over a  broad range of collision energies, while the ortho- results yield clearly smaller cross sections as the collision energy increases to about 50 cm $^{-1}$. The purely vibrationally inelastic data in the uppermost left panel indicate, for this particular case of  state-changing dynamics, that the cross sections are nearly coincident with each other, with the para values still exhibiting a more marked presence of resonant peaks.

In the Section below we shall further analyse the effect of rotational excitation in either ortho- or para-H$_2$  on the vibrational cooling in C$_2^-$ by looking at the behaviour of the computed rate coefficients.

\subsection{Computed ro-vibrationally inelastic rate coefficients}
\label{sec:vib_rates}

The  inelastic cross sections of the previous Section can be used to obtain the corresponding thermal rate coefficients, which can be evaluated as the convolution of the computed inelastic cross sections over a Boltzmann 
distribution of the relative collision energies of the interacting partners as

\begin{widetext}
\begin{equation}
k_{{\nu j_1j_2 \rightarrow \nu^{'}j_1^{'}j_2^{'}}}(T) = \left(\displaystyle \frac{8}{\pi \mu k_{B}^3 T^3 } \right)^{1/2} 
\times
\int_0^{\infty}E_{\textnormal{coll}} \sigma_{{\nu j_1j_2 \rightarrow \nu ^{'}j_1^{'}j_2^{'}}}(E_{\textnormal{coll}}) e^{-E_{\textnormal{coll}}/k_{B}T}dE_{\textnormal{coll}}
\label{eq.rateK}
\end{equation}    
\end{widetext}

As discussed in the Introduction Section, studies on laser cooling of C$_2^-$ have assumed that the anion was initially cooled to tens of Kelvin
\cite{15YzHaGe} and thus the rate coefficients were computed between 5 K and 100 K at 1 K intervals. The important contribution of our present detailed calculations will be to assess the individual probabilities of populating by a collision a variety of rotational states of the anion along with the main vibrational cooling process. The experiments we described in reference \cite{23NWL}, in fact, were not observing which rotational states will be populated after the anion's decay to its $\nu = 0$ level in the trap with the H$_2$ partner, so the computational estimates will be important to know.

\begin{figure}
\centering

\label{fig9a}
	\includegraphics[width=0.93\linewidth,angle=+0.0]{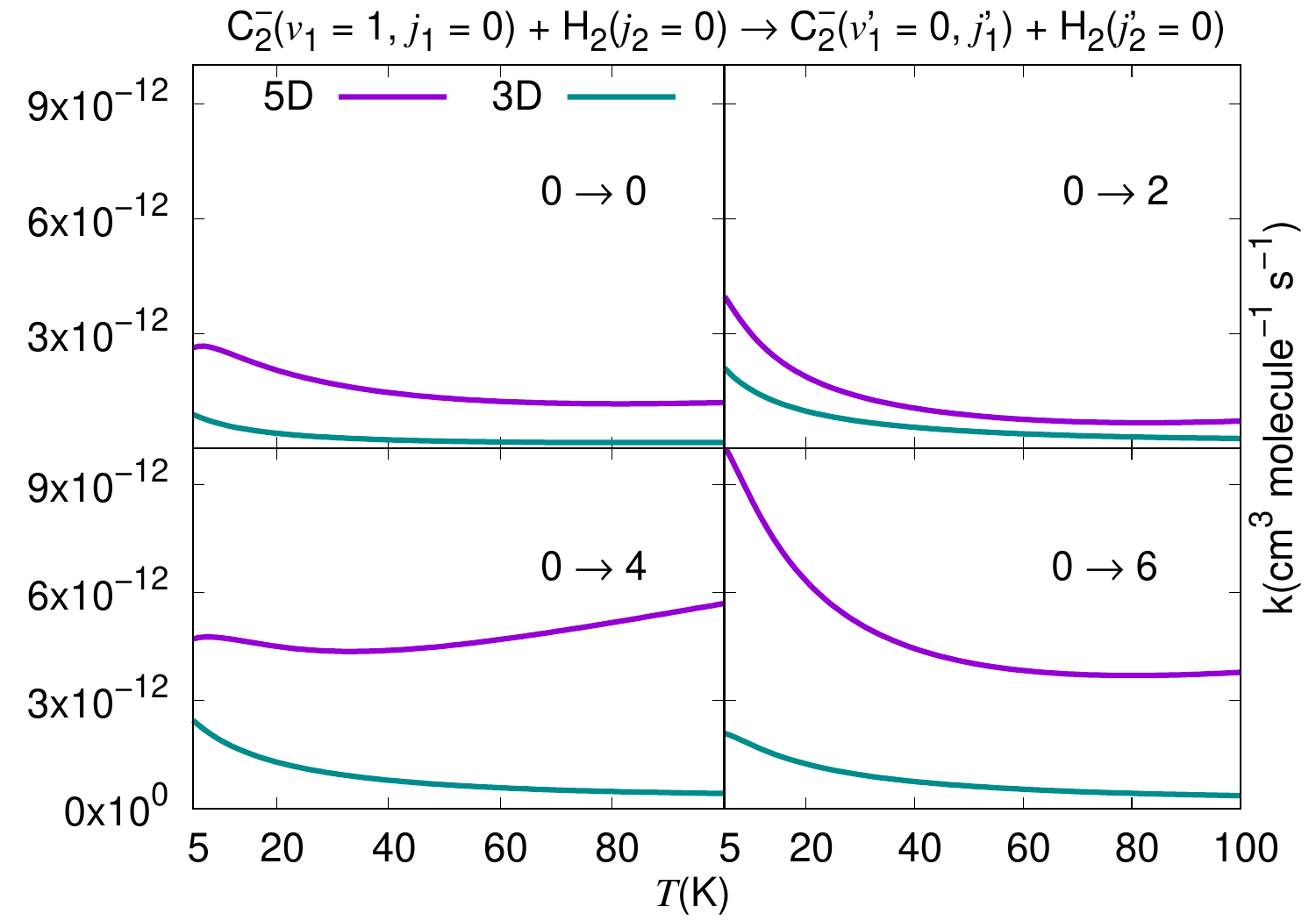}

\label{fig9b}
	\includegraphics[width=0.93\linewidth,angle=+0.0]{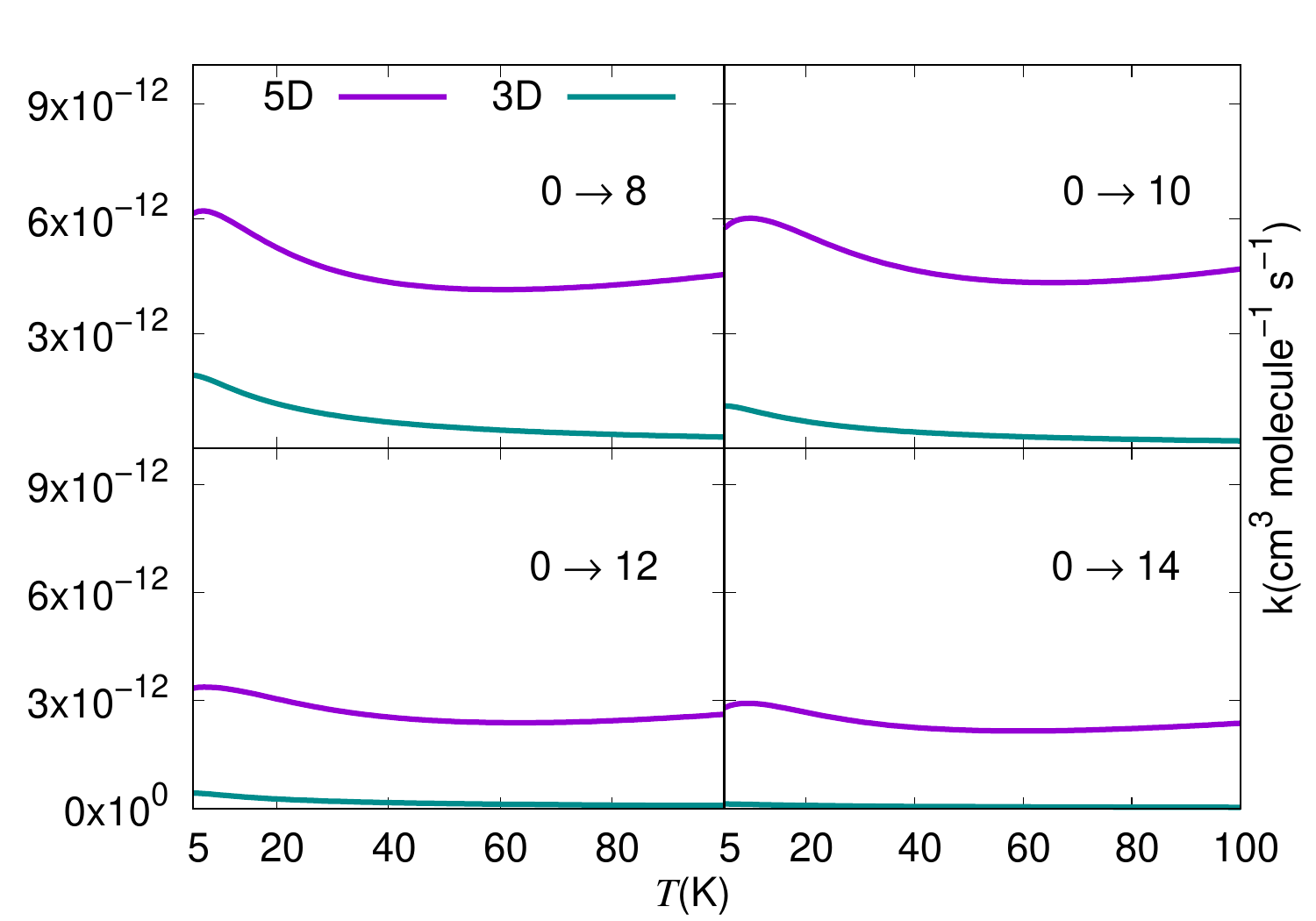}

\caption{Computed vibrational cooling rate coefficients of the anion from the $\nu = 1$ to the $\nu = 0$ level, with concurrent rotational excitation processes as listed in each panel. Full calculations using the 5D interaction are given by purple lines, while the earlier results using the 3D interaction are given by green lines. Both sets of calculations have the para-H$_2$ as the collision partner. The upper set of panels report transitions to the lower rotational levels of the anion, while the lower set of panels shows transitions into higher rotational states of the anion. See main text for further details.}
\label{fig9}

\end{figure}

To verify once more the effect of using the 5D PES to carry out the relevant inelastic dynamics, we report in the two sets of four panels of Figure 9 a comparison of the inelastic rate coefficients obtained using the full dynamics (purple lines) and those we had computed in our earlier work \cite{23NWL} for the same range of inelastic processes. All processes are obtained using the para-H$_2$ as collision partner. The rotational excitation processes involve only states of the anionic partner and correspond to the pure vibrationally inelastic process in the uppermost left panel, while all following panels report final populations of higher rotational states of the vibrationally quenched anionic molecule. 

We see in the  upper set of  panels, which correspond to either no energy being transferred to rotational levels or to the smallest amount of it being transferred to rotation, that the results using the 3D PES (green curves) are very close to the purple curves obtained via the 5D PES at lower temperatures. As more energy is given to rotational excitation the two sets of calculations differ much more from each other in size and in their dependence on temperature. Furthermore, the larger the amount of energy that goes into rotation larger are the corresponding rate coefficients, a feature of the present findings which we shall further analyse below.

The calculations reported by the lower four panels of Figure 9 analyse the same type of inelastic processes  while however considering rotational excitation of the vibrationally quenched anionic partner to higher rotational states: from the $j_1'$ = 8 excitation on the upper left panel, to the excitation into $j_1'$ = 10, 12 and 14 in the remaining three panels. In all instances we see that the two sets of rates are more different from each other, indicating that the data from the purple curves are larger by a factor of ten  than those given by the green curves. In all panels the results of the purple curves are larger than those  for purely vibrational transfer shown in the uppermost-left panel of Figure 9.

Initially, this may appear counter intuitive. However, one realises that the overall amount of energy being transferred away from the molecule diminishes  as the higher rotational states of the vibrationally quenched ($\nu'$ = 0) state of C$_2^-$ becomes populated.  While the 0$\rightarrow$0 transition corresponds to a loss of 1754 cm$^{-1}$, the successive higher $j_1'$ transitions (0$\rightarrow$2, 0$\rightarrow$4, ,...0$\rightarrow$14) involve an increasing amount of energy being kept as internal rotational energy, with the actual energy transferred away becoming  less (1754 - 1.729 ($j_1'$ ($j_1'$ + 1)) and the degeneracy of the $j_1'$ state increasing as (2$j_1'$ + 1). Hence, we could see that the  reduction of the amount of energy being dissipated causes a greater efficiency for the internal energy redistribution after collisions. Such trends turn out not to change much over the examined range of temperatures.

\begin{figure}[h!]
\centering
\label{fig10a}
	\includegraphics[width=0.90\linewidth,angle=+0.0]{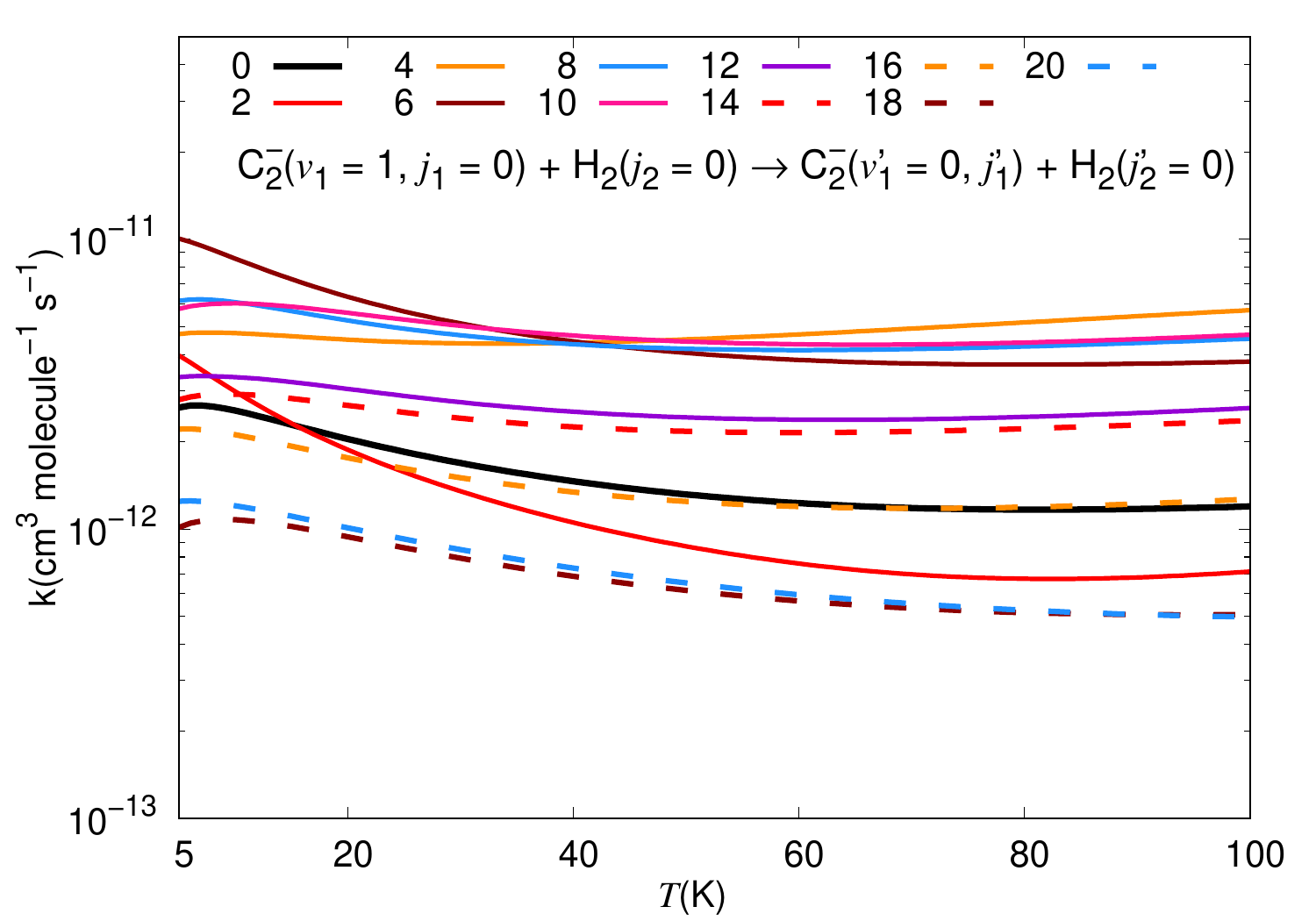}
 
\label{fig10b}
	\includegraphics[width=0.90\linewidth,angle=+0.0]{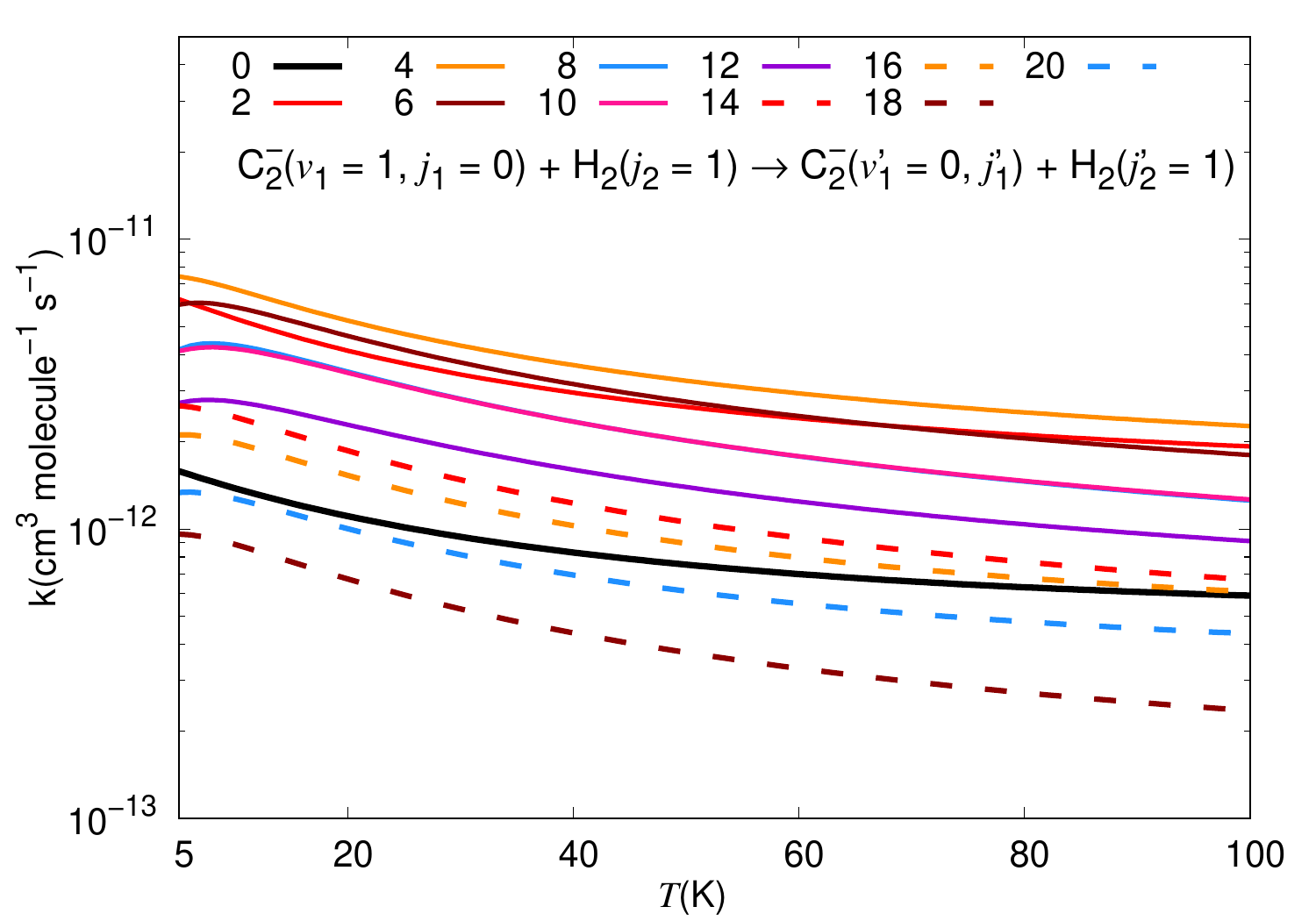}
 
\caption{The upper panel reports (5D) computed  rate coefficients for inelastic processes involving only the molecular anion and with the para-H$_2$ as a partner, while the lower panel reports the  same processes as above, but with the ortho-H$_2$ as a neutral  partner. See the main text for further comments.}
\label{fig10}
\end{figure}

By looking at the results given in the two panels of Figure 10, we realize that to have either para-H$_2$ or ortho-H$_2$ as a collision partner changes the size of the involved rates only marginally. They turn out, however,  to be close to each other at the lowest collision energies, with the ortho-results becoming somewhat smaller as the collision energy increases. On the whole, however, we see a rather minor effect coming from either nuclear spin arrangements as a collision partner.

\begin{figure}[h!]
\centering
\label{fig11a}
	\includegraphics[width=0.85\linewidth,angle=+0.0]{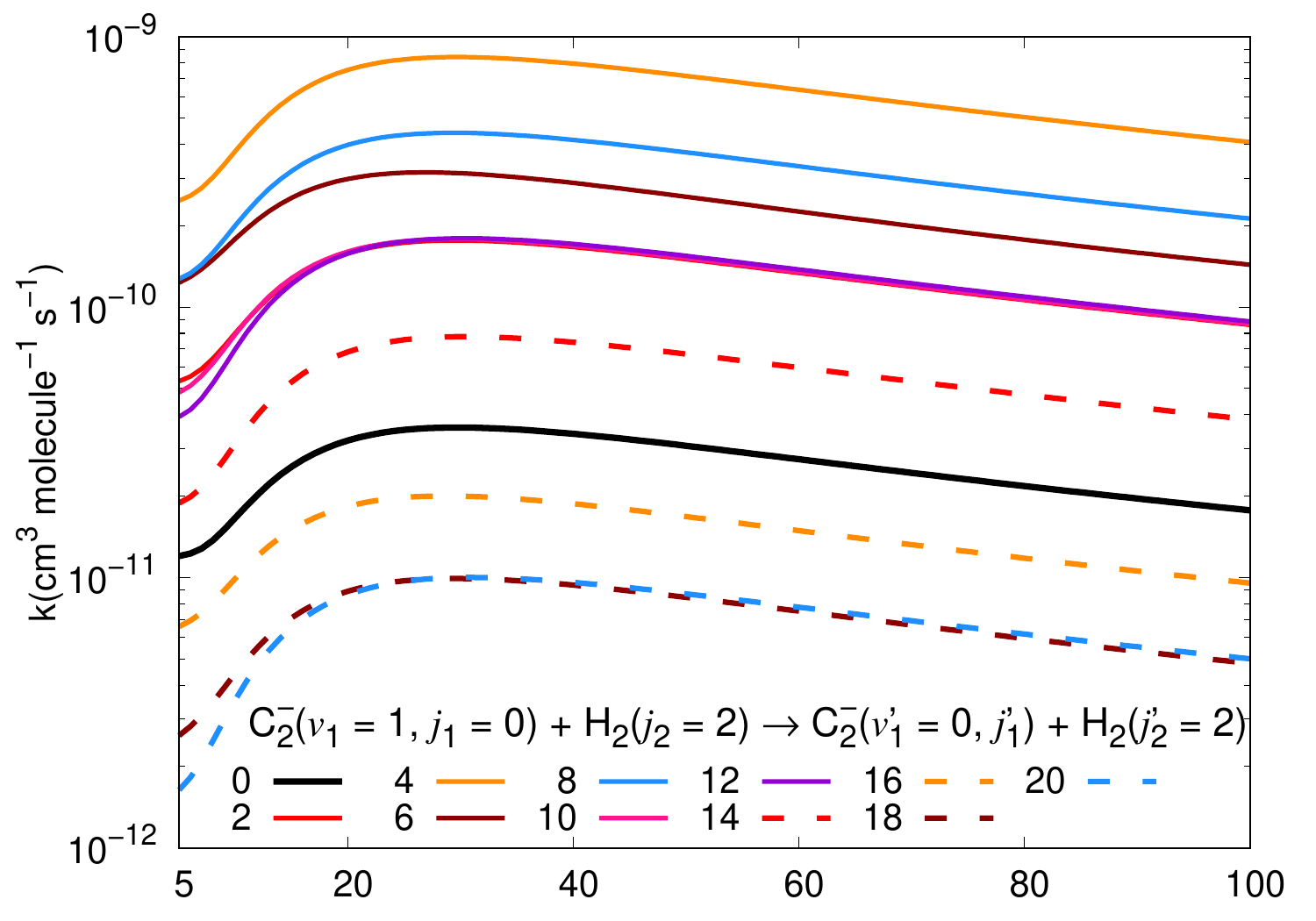}
 
\label{fig11b}
	\includegraphics[width=0.85\linewidth,angle=+0.0]{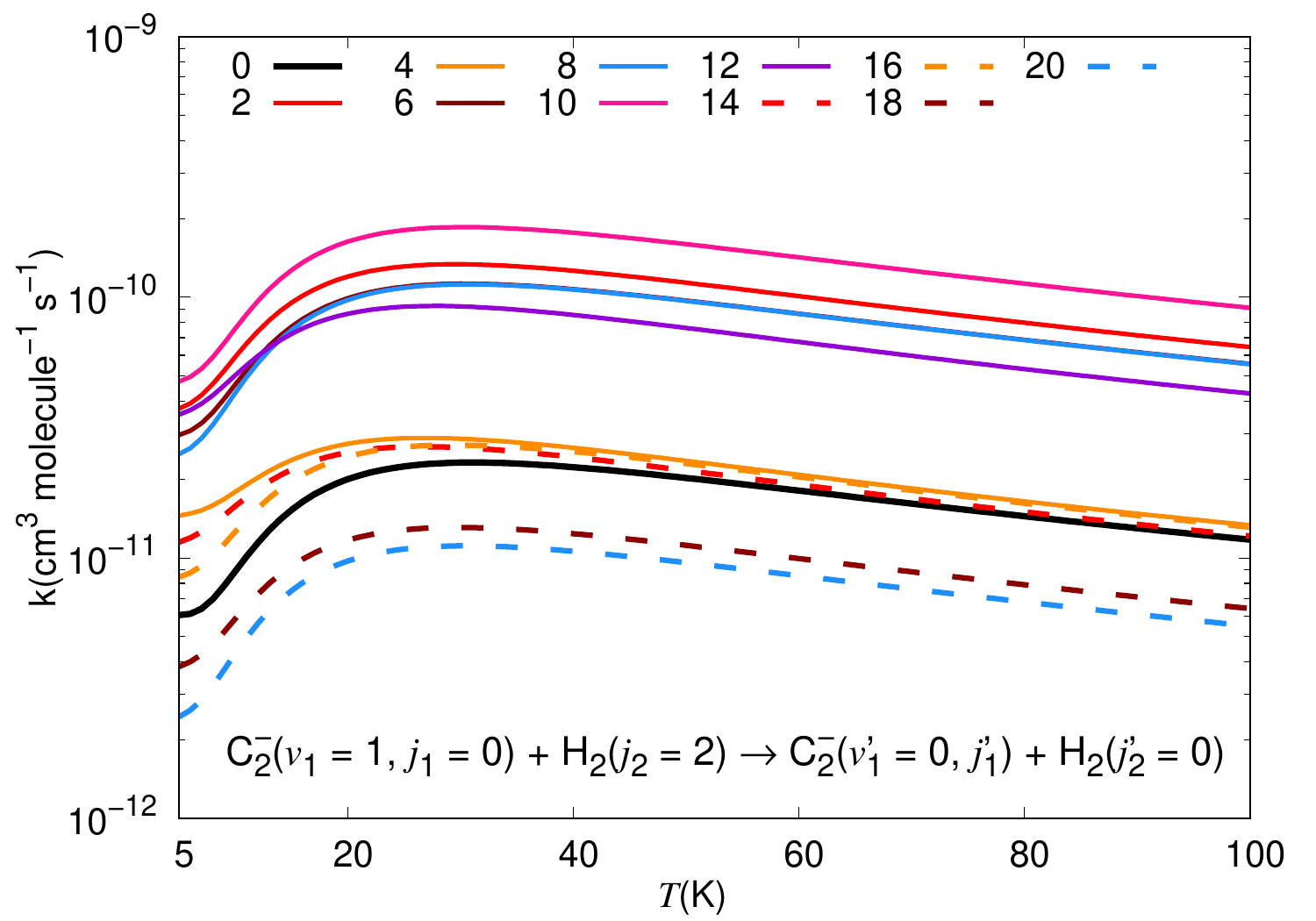}

\caption{The upper panel reports (5D) computed rate coefficients for inelastic processes involving  the molecular anion, with the rotational states changing  up to $j_1'$ = 20, while the para-H$_2$ partner remains excited in its $j_2$ = 2 rotational state. The lower panel reports the same processes as above but  including the rotational de-excitation of the para-H$_2$ partner. See the main text for further comments.}
\label{fig11}
\end{figure}

The calculated rate coefficients reported by the panels of Figure 11 and Figure 12 are considering  more complex situations whereby the excitation of rotations involves not only the anionic partner but also either the para-H$_2$ or the ortho-H$_2$ molecular rotor. Such analysis will allow us to evaluate the relative energy transfer efficiencies between the two partners when both undergo rotational  state-changes during the collision, the latter being chiefly driven by the vibrational cooling of the C$_2^-$ partner. Only collisions using the 5D quantum dynamics are shown in the two panels of each of the two Figures.

\begin{figure}[h!]
\centering
\label{fig12a}

\includegraphics[width=0.85\linewidth,angle=+0.0]{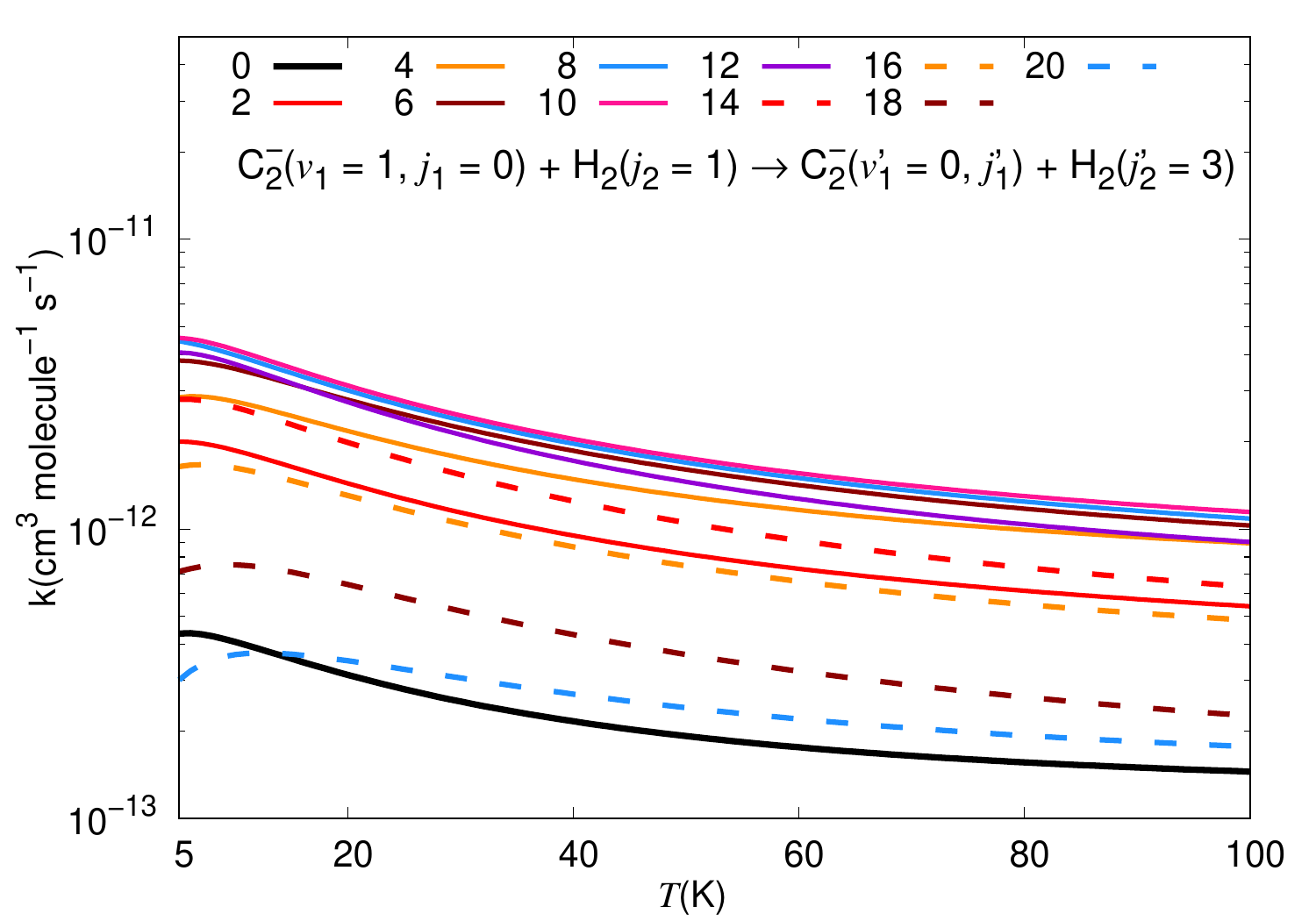}
 
\label{fig12b}
	\includegraphics[width=0.90\linewidth,angle=+0.0]{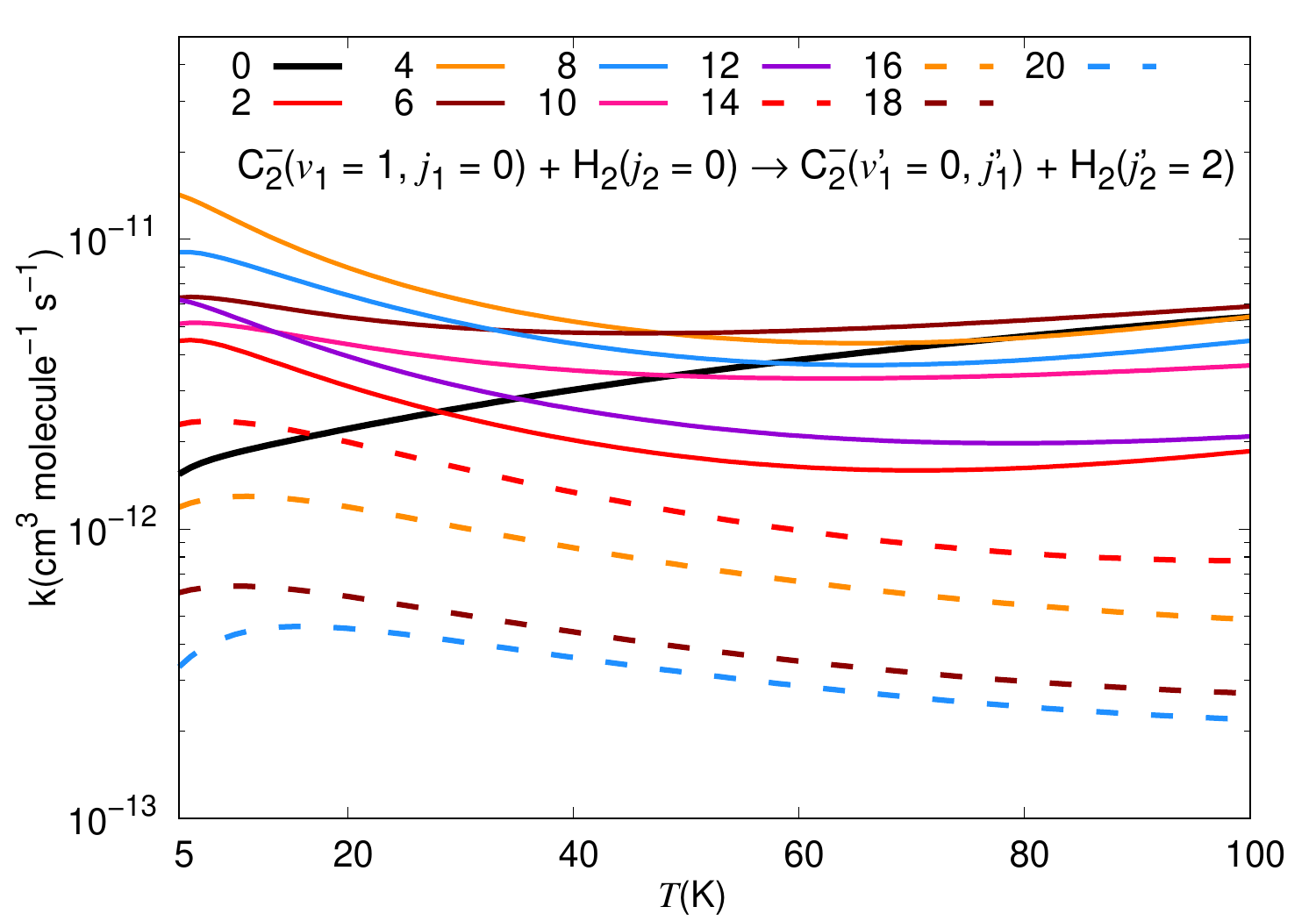}
 
\caption{The upper panel reports (5D) computed  rate coefficients for inelastic processes involving  the molecular anion with the concurrent rotational excitation of the ortho-H$_2$ as a partner, while the lower panel reports the values for the same processes as above, but including the rotational excitation of the para-H$_2$ partner. See the main text for further comments.}
\label{fig12}
\end{figure}

\begin{figure}[h!]
\centering
\label{fig13a}

\includegraphics[width=0.90\linewidth,angle=+0.0]{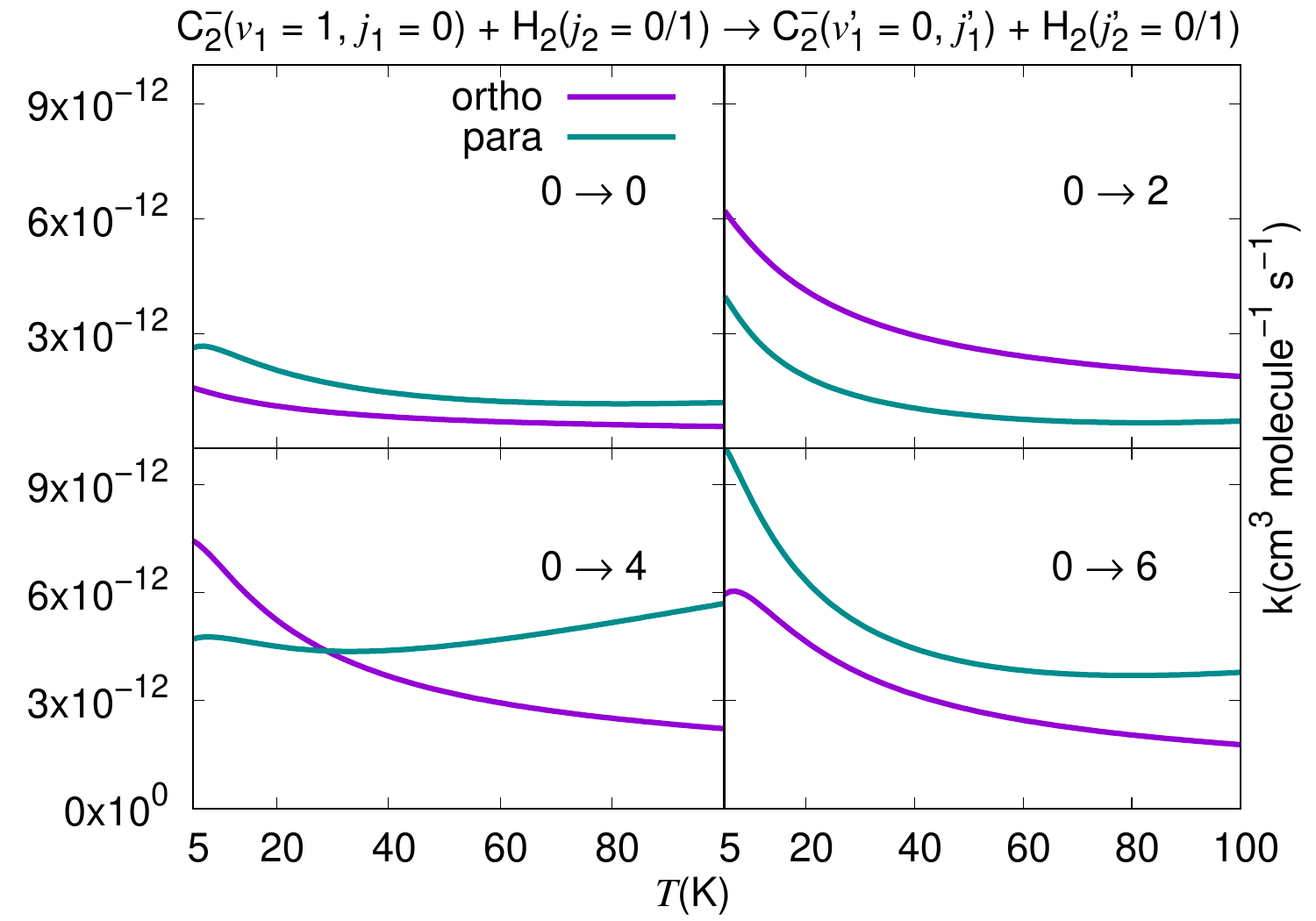}
 
\label{fig13b}
	\includegraphics[width=0.90\linewidth,angle=+0.0]{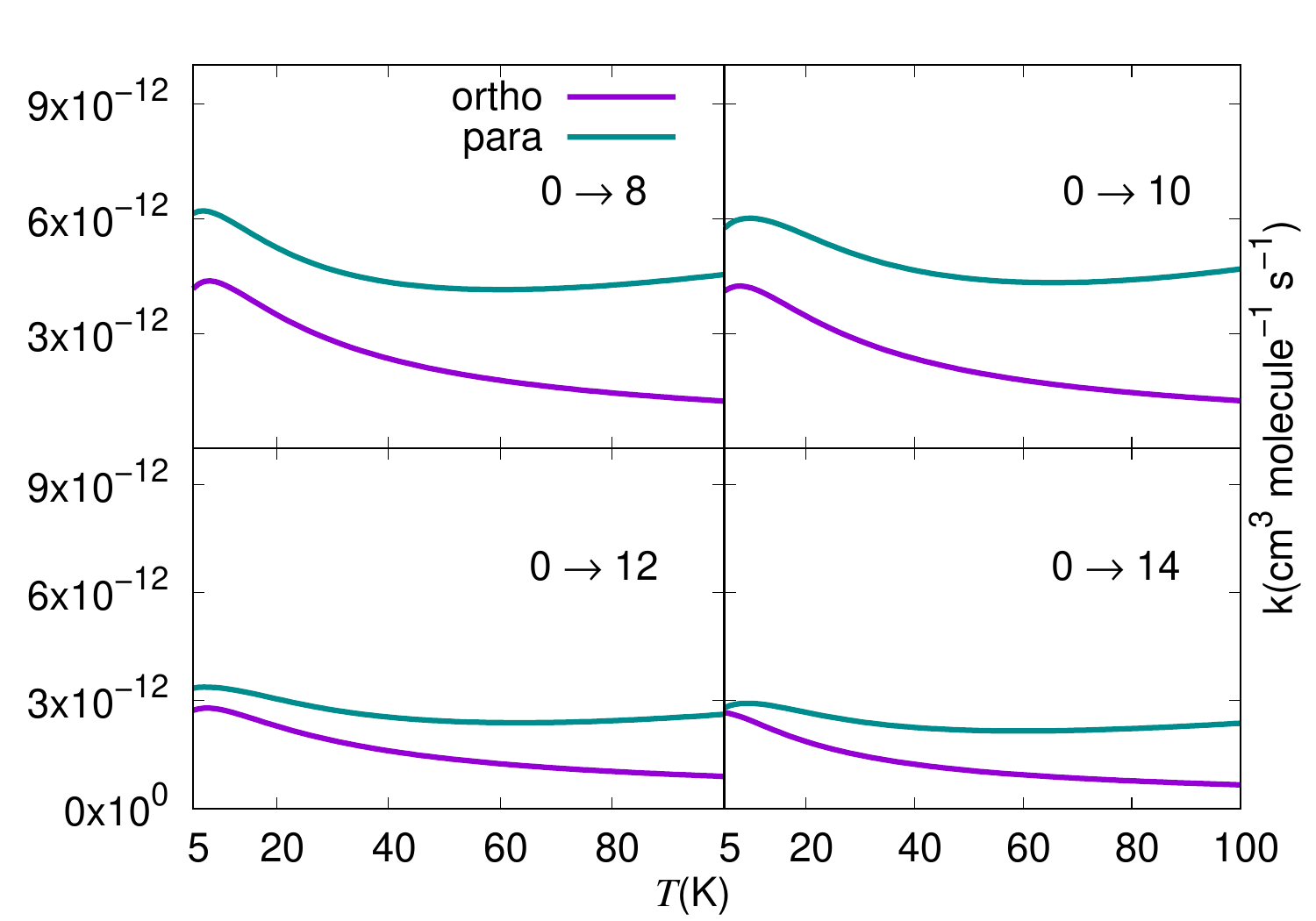}
 
\caption{The two panels report (5D) computed  rate coefficients for inelastic processes involving  the molecular anion colliding with either the ortho-H$_2$ or  the para-H$_2$ partner in their ground rotational states. The different sub-panels indicate different rotational states of the anion which are being populated during each collision process.  See the main text for further comments.}
\label{fig13}
\end{figure}

The upper panel of Figure 11 shows rotational excitations of rotor states of the  C$_2^-$ partner from $j_1'$ = 0 up to $j_1'$ = 20, with  the para-H$_2$ partner in its excited $j_2'$ = 2. The relative probabilities for the various excitation processes are clearly changing when the para-H$_2$ partner is excited and turn out to be markedly larger than those obtained with the partner molecule is in its ground rotational state (upper panel of Figure 11). On the other hand, the results in the lower panel of the same figure indicate that the concurrent de-excitation of the para-H$_2$ partner causes all transition probabilities to become somewhat smaller, although still larger than those given by the upper panel of Figure 11. We see once more that the size of the various processes is largely controlled by the dominant size of the initial vibrational de-excitation process, so all rate coefficients never get larger than about 10$^{-9}$cm$^{3}$molecule$^{-1}$s$^{-1}$ at the most, while being largely around 10$^{-10}$cm$^{3}$molecule$^{-1}$s$^{-1}$.

In the two panels of Figure 12 we further present the cases of rotational excitation of the various states of  C$_2^-$ up to $j_1'$ = 20, with the concurrent excitations of either the ortho-H$_2$ partner (upper panel) or the para-H$_2$ partner (lower panel). We clearly see from the data of both panels that to involve the much larger energy of the H$_2$ rotors generates rate coefficients which are  smaller than in the case where only the anionic partner undergoes rotational excitation (see the panels of Figure 12). In the lower temperature range, in fact, their sizes are down by nearly one or two orders of magnitude with respect to those in Figure 12, and remain so even at the  highest temperature values we have considered. This feature of the rate coefficient behaviour will be further analysed below.

An additional comparison of the various rotational final states of the anion which become populated during the vibrational cooling process is reported by the data of Figure 13. We indicate there the cases where the neutral molecular partner is always in its ground rotational state and is given by either the para-H$_2$ (green lines) or by ortho-H$_2$ (purple lines). The various sub-panels further report, as before, different final rotational states of the anion being populated during the collisions. We can make the following comments about the shown data in all the Figure panels:

(i) for the purely vibrational de-excitation process (shown in the uppermost left panel) we see that the rates  with the para-H$_2$ partner are larger over the whole temperature range considered;

(ii) when the lower rotational states of the anion are also populated (upper four sub-panels) we see that the ortho-rate coefficients are larger only for the case where the (0 $\rightarrow$ 2) rotational excitation process occurs, while constantly remaining smaller by varying factors between 2 and 5 over the results of also the lower four sub-panels;

(iii) We can therefore say that the present calculations involving state-to-state transitions of the ro-vibrational states of the anion yield smaller probabilities at the temperature of interest when the ortho-H$_2$ is considered as a collision partner.

Such findings are expected to produce significant differences when comparing the present calculations with existing experiments, as we shall do in the following sub-section.

\subsection{Comparing the 5D computed rate coefficients with  the available experimental results}

Since we wish to further compare the five-dimensional  vibrational quenching rate coefficients calculated in the present work  to those obtained in the experiments, we have to look more closely to the conditions in which observations have been made in our earlier work of Ref. \cite{23NWL}. 

First of all, following the information from the experimental conditions, we shall assume that the excited vibrational  state with $\nu$ = 1 of the ground electronic state of C$_2^-$, which has been prepared by laser cycles via the excited electronic states (as described in the Introduction), is reaching a Boltzmann distribution for the initial rotational states at the temperature of the trap conditions. This is justified by the fact that rotational quenching is known to be far faster than vibrational quenching, so that we can learn from experiments which rotational states of the initial vibrationally  excited anion, will be populated during the ensuing collisional quenching step, while however we cannot say which will be the populated rotational states of the quenched anion after collision.

The observable vibrational quenching rate constant $k_{\nu \to \nu'}(T)$ could thus be modeled
 from our calculations by summing over all final rotational states, reached after the cooling down of the C$_2^-$ to the  $\nu$ = 0 state by collision, for which we find the quenching rate coefficients to be significant in size. On the other hand, we are able to carry out a Boltzmann averaging over all the  initial rotational 
states of the anion, in the $\nu$ = 1 state, which are known to be populated  before the collisional quenching. The temperature of the latter set of  C$_2^-$ molecules was considered in the experiments of \cite{23NWL} to be close to 20K .

Hence, we can write the following:
\begin{widetext}
\begin{eqnarray}
k_{\nu \to \nu'}(T) = \frac{1}{\sum_{j_1} (2j_1 + 1) \exp(-\epsilon_{\nu j} - \epsilon_{\nu = 0 j_1 = 0} ) /k_{B}T)} \nonumber 
\times
(\sum_{j_1}(2j_1 +1) \exp(-\epsilon_{\nu j} - \epsilon_{\nu = 0 j_1 = 0}) /k_{B}T)   \nonumber
\times
\sum_{j_1'}  k_{\nu j \to \nu'=0 j_1'}(T) )
\hspace*{3.5cm}
\label{eq.rateK_av}
\end{eqnarray}
\end{widetext}

To simplify the notation in the above equation we have replaced the subscript ${(\nu,j_1,j_2)}$ with the simpler form ${(\nu,j)}$ .

We actually have covered a broader range of temperatures of interest (5 to 100 K) in our calculations. As discussed in previous Sections we have treated the C$_2^-$ anion as a pseudo-singlet since we have verified in many earlier studies of our group (e.g. see: \cite{20MaGiWe} and ref. quoted therein) that the effect of actually including the fine structure terms alters the final cross section values by less than a few percent.  We also know that only the first few rotational states are significantly populated at the temperature of the carried out experiment (e.g.: $j_1$ = 0 to $j_1$ = 4) and so only these were used as starting $j_1$ states. On the other hand, all the final $j_1'$ states used in our basis set, and producing significant rate coefficient values, went up to
 $j_1'$ = 20, so that all the calculations into those final states were used in the sum indicated in eq.(16).

It is instructive at this point to pictorially verify the relative size distributions of the final states ending up with the $\nu$ = 0,  but with different $j_1'$, by looking at a 'bar spectra' presentation of the computed data. The panels of Figure 14 report such data for three different temperature values and for three different choices of $j_1$ when the anions are in the $\nu$ = 1 state before the collisional quenching of the vibration. Only the case where the para-H$_2$ is considered as collision partner are shown in that Figure.

\begin{figure}
\centering

\label{fig14a}
	\includegraphics[width=0.97\linewidth,angle=+0.0]{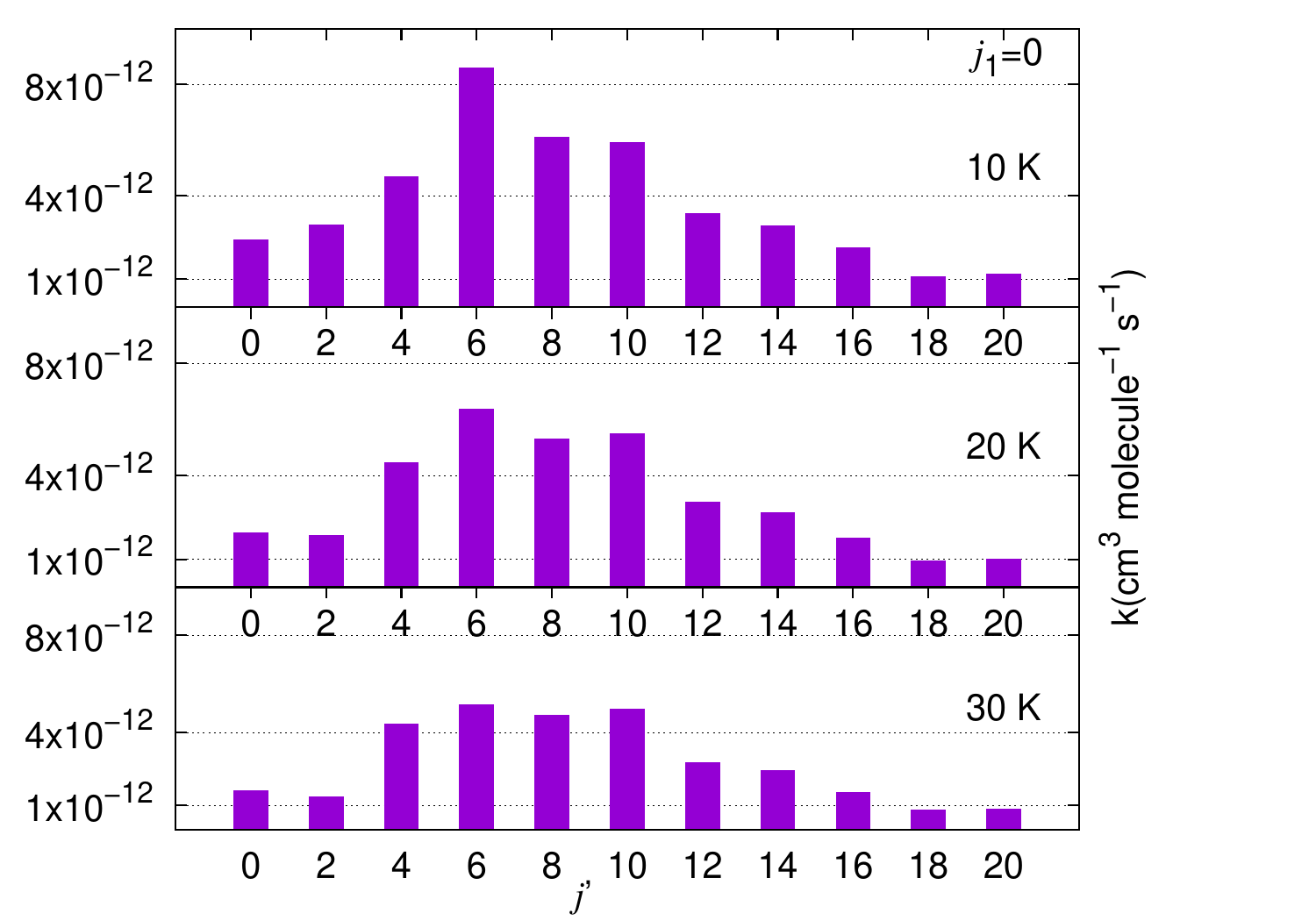}

\label{fig14b}
	\includegraphics[width=0.97\linewidth,angle=+0.0]{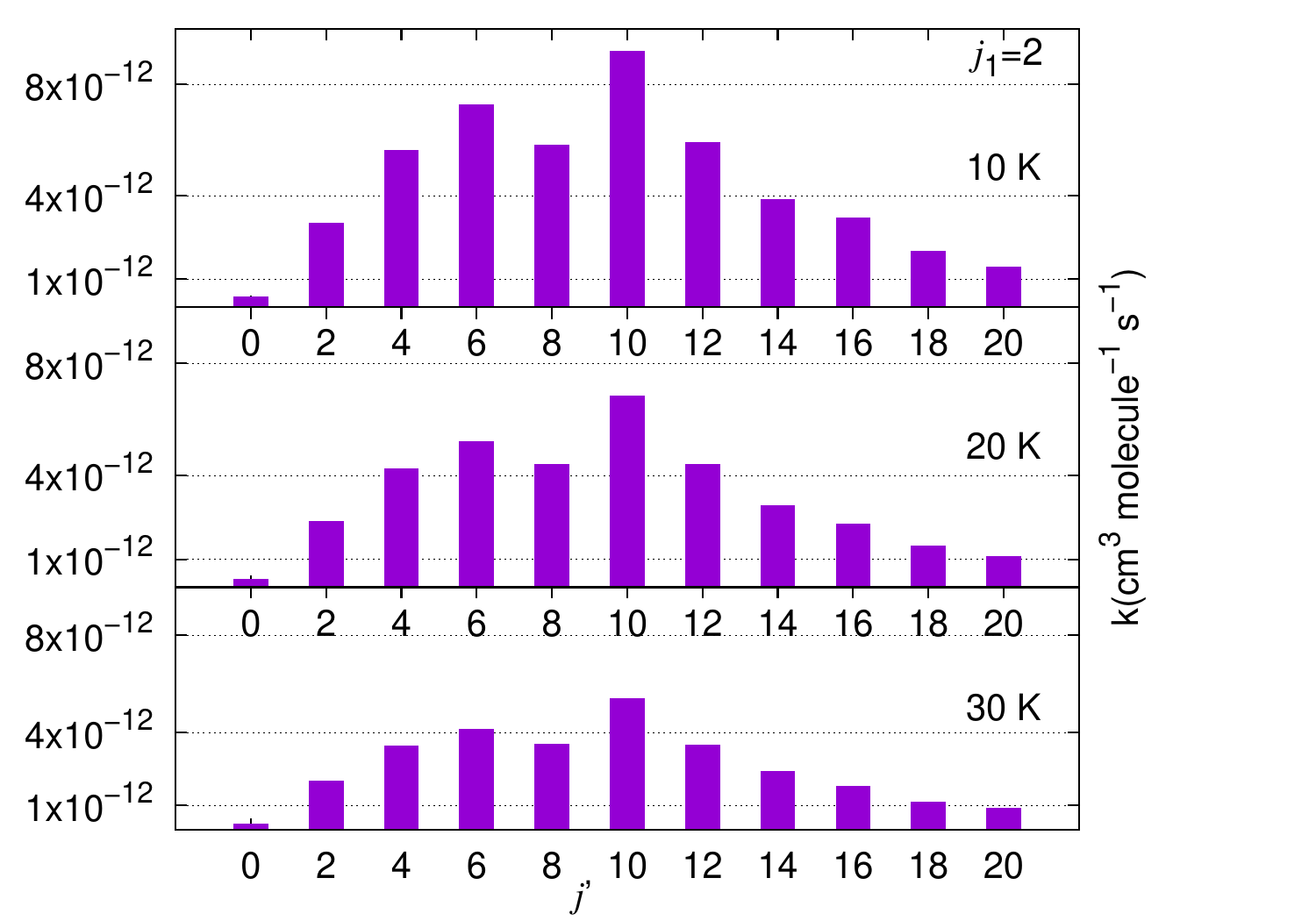}

\label{fig14c}

	\includegraphics[width=0.97\linewidth,angle=+0.0]{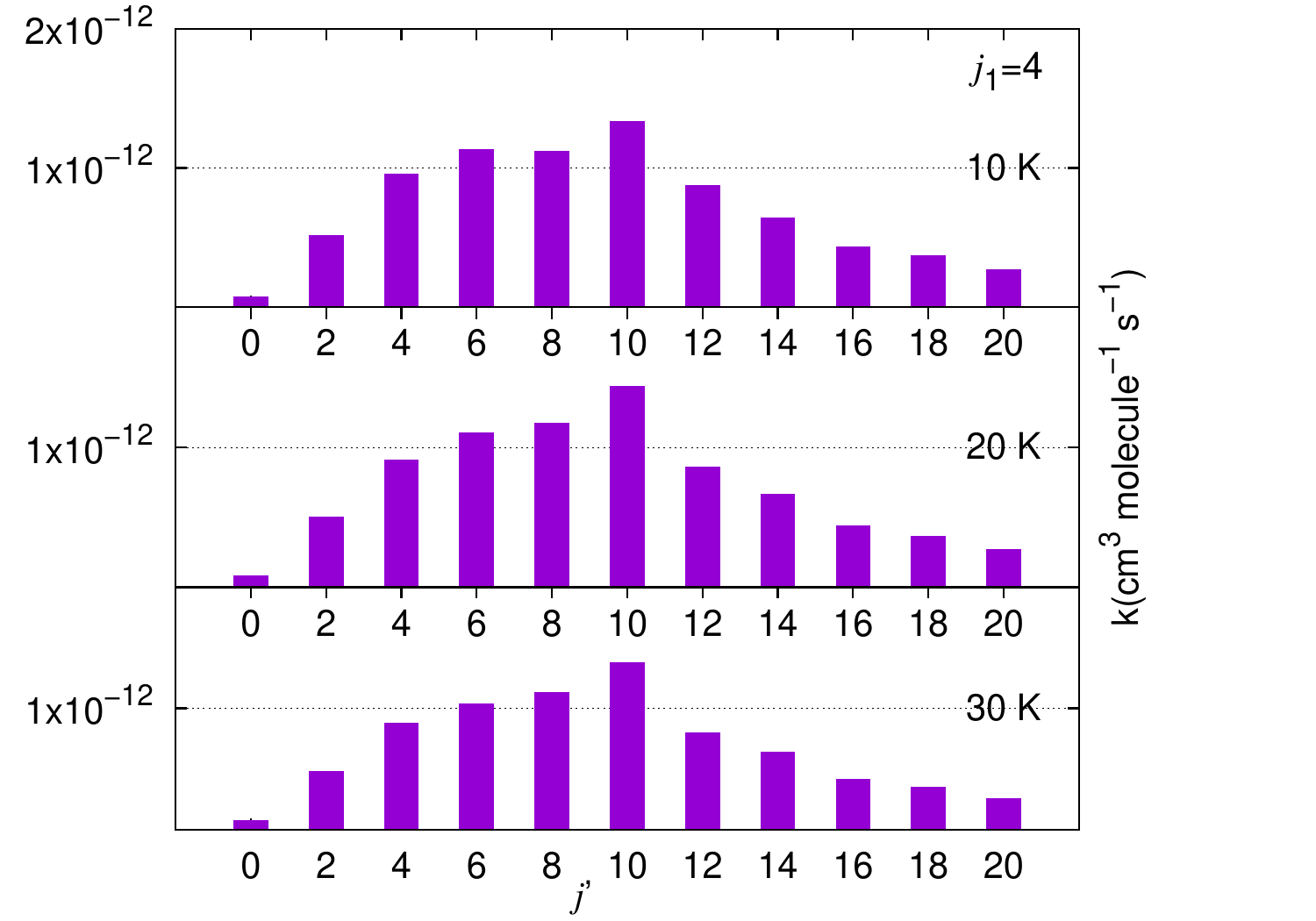}

 \caption{Computed rate coefficient distributions over the final rotational state after collisional cooling of the anion with para-H$_2$($j_2$=0) as partner. The data are  presented as vertical 'bars' for each final rotational state. The initial state before the vibrational cooling changes in each set of panels: upper panel  has  $j_1$ = 0 as the initial state; the middle panel has $j_1$ = 2 and the lower panel has $j_1$ = 4. Three different temperatures are shown in each of the sub-panels in each main panel. See the main text for further details.}
 
\label{fig14}

\end{figure}

The following observations could be made when looking at the rate coefficient distributions shown in the various panels of Figure 14:

(i) when the initial rotational state  $j_1$ = 0 the probability of transferring energy into excited rotational states peaks at $j_1'$= 6 for all temperatures, with the rate coefficients for $j_1'$ = 8 and 10 also being significantly large. The higher final states exhibit successively smaller probabilities of occurrence;

(ii) when the initial states are either $j_1$ = 2 or 4, the largest rotational energy transfers are seen for $j_1'$ = 10, hence showing a similar tendency as in the previous case;

(iii) all the final rate coefficient values within each distribution consistently show that they become smaller when the initial state is  $j_1$ = 4. 

(iv) in addition, we see from all the data in all the sub-panels in Figure 14 that the general trend as the final rotational state becomes larger is that the corresponding rate coefficients get larger, but up to a maximum j$_1'$ after which the rate coefficients become smaller. These findings confirm that the quantum dynamics of angular-momentum transfer probabilities gets increasingly less efficient when larger angular momentum changes occur during the collision, while also indicating that the smaller values of the rate coefficient pertain to the final states with the j$_1$' =0.

The above findings certainly bear importance when we shall analyse  the existing experimental datum taken from \cite{23NWL} and reported with the present computed findings by the curves given by Figure 15.

\begin{figure}[h!]
\centering
\includegraphics[scale=0.37,angle=+0.0,origin=c]{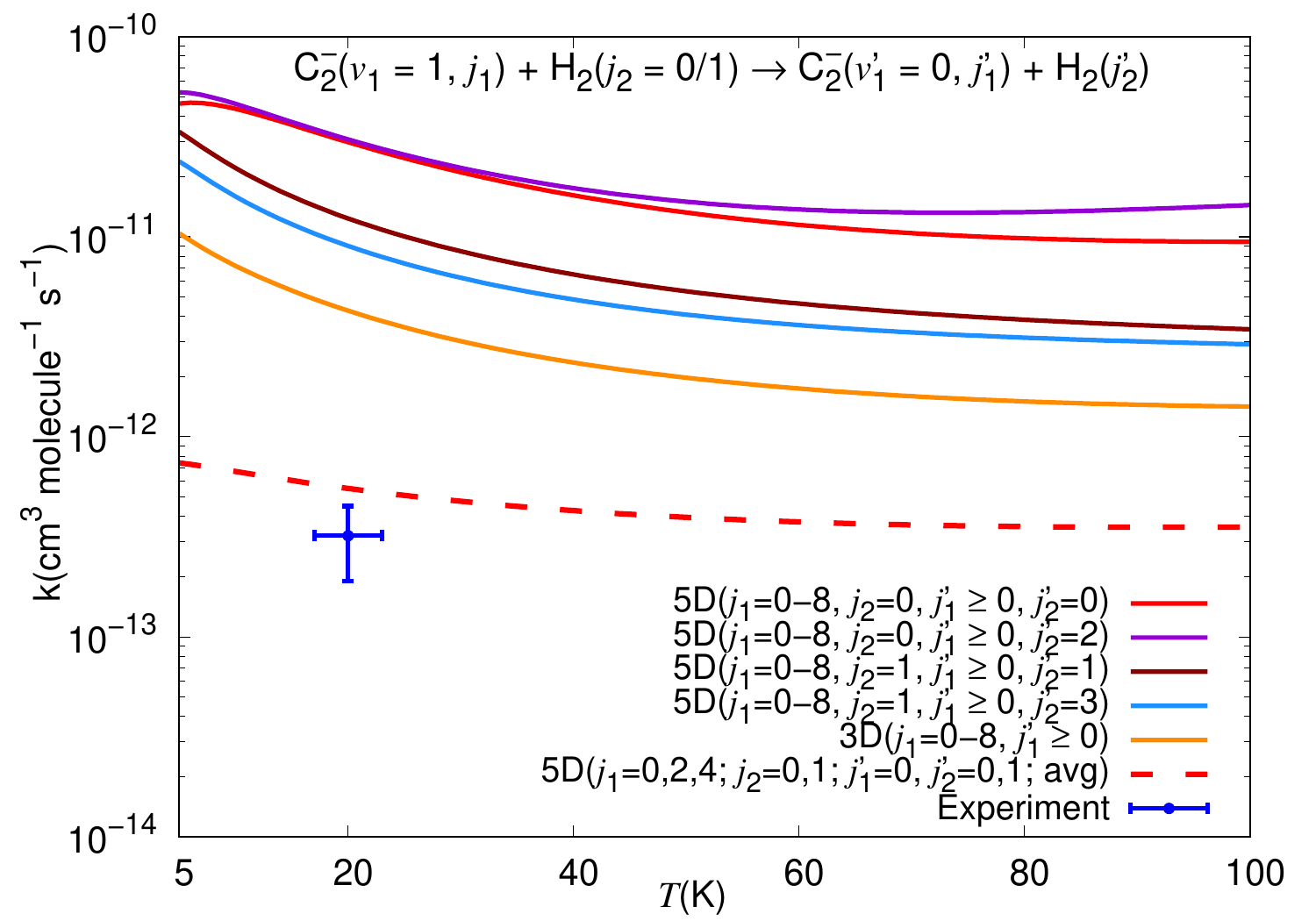}

\caption{Computed rate coefficients for vibrational de-excitation for the anionic partner with all the rotationally inelastic processes Boltzmann averaged at $T$ = 10-100 K. See the main text for details. The purple curve refers to 5D dynamics where the para-H$_2$ partner is also rotationally excited from its ground state to the $j_2'$ = 2 state, while the red curve only involves rotational states of the anion. The two lower curves (brown and blue) describe the same set of processes but this time having ortho-H$_2$ as a partner. The yellow curve is like the red curve but obtained using the 3D dynamics reported in our earlier work \cite{23NWL}.All the initial states used to obtain these curves have been thermally averaged over rotational (from 0-to-8)  initial states according to eq.(16). The dashed red curve reports rate coefficient values obtained by averaging the dominant first three initial $j_1$ states with the experimental distribution at 20K  but only including the final rotational state with $j_1'$ = 0. It is further obtained by the weighted sum ( 3/4 + 1/4) of the results involving either the ortho- or the para-H$_2$ collision partner. The single blue point corresponds to the only available experimental observation from reference \cite{23NWL}. }
\label{fig: Figure15}
\end{figure}

The results of our present calculations,  reported by the curves shown in  Figure 15, are now  modeling the expected distributions into the final rotational states of the  collisionally cooled anion, as already discussed in the previous paragraphs. More specifically, the red curve there presents  the averaged rate coefficients  obtained following eq.(16) and involving the concurrent rotational excitation of the para-H$_2$ partner, while the purple curve reports the same calculations from eq.(16) but only involves the population of the rotational  states of the anion. The next two curves given by solid lines (brown and blue lines) show the same processes given by the two upper curves but dealing with the ortho-H$_2$ as a partner: they clearly show that the rate coefficients linked to the ortho-H$_2$ as partner are  smaller than when the collision partner is the para-H$_2$ molecule. The slightly lower  yellow curve presents the earlier 3D  calculations of our earlier work \cite{23NWL}, where only excitation of the anion's rotations were considered in the inelastic process and the para-H$_2$ was the only possible collision partner, as we have discussed in the earlier Sections.

The only existent experimental point estimated to be at 20 K, is given in the Figure by the blue cross and taken from \cite{23NWL}. It indicates that all full 5D calculations yield invariably larger rate coefficients than the one given by the experiment, even when the excitation of either the para- or the ortho-H$_2$ partner is included in the quantum dynamics. The calculations where the more realistic 5D PES is employed within the quantum dynamics invariably yield larger values which are nearly two orders of magnitude larger than the available  experimental point.

To possibly better understand the findings of the present comparison, we have added in Figure 15 one additional set of results, given by the dashed red curve. The dashes report purely vibrational cooling, starting from the thermal averaging of the  three dominant rotational states of the initial anion before the collisional cooling: $j_1$ = 0, 2 and 4 at 20K. On the other hand, only the final, vibrationally  cooled anionic states with $j'_1$ = 0 are considered. The calculations are those obtained using the present extended surface in 5D and only for that single process. Furthermore, we have considered the role of either the para- or the ortho-H$_2$ partner in collision with the present anion and have included in the data for the red, dashed curve a weighted sum of (1/4 + 3/4) of the two sets of processes.

We see now that the purely vibrationally inelastic processes which occur  without formation of rotationally excited anions in the $\nu$ = 0 state get markedly closer to the existing experimental point. The efficiency of the 5D calculations, at the experimental temperature of about 20 K, makes the initially averaged 5D calculations to be about 5.0 x 10$^{-13}$cm$^3$molecule$^{-1}$s$^{-1}$, to be compared with an experimental value of about 4.0 x 10$^{-13}$cm$^3$molecule$^{-1}$s$^{-1}$. Such results are clearly very close to each other.

Another interesting effect is provided by the analysis of the size and behaviour of the inelastic rate coefficients shown by Figure 15 when we separately  analyse the collision results dealing with either the ortho-H$_2$ or the para-H$_2$ partner of the main anion. This analysis is reported by the data of Figure 16.
These data clearly indicate the following features:

(i) rotational excitation in the para-H$_2$ partner produces by far the largest vibrational cooling rate coefficients;

(ii) in contrast, rotational excitation in ortho-H$_2$ as a collision partner leads to the smallest vibrational cooling rates, well below the experimental finding;

(iii) the presence of either pure ortho- or pure para-H$_2$ partners (red and brown curves) without them being excited has a much smaller effect on the vibrational cooling rates;

(iv) when, however, one considers the relative weights of the two species in normal-H$_2$ (dashed red and black curves) we see that the difference in cooling efficiency becomes again rather small and the two dashed curves are close to the experimental value.

\begin{figure}[h!]
\centering
\includegraphics[scale=0.37,angle=+0.0,origin=c]{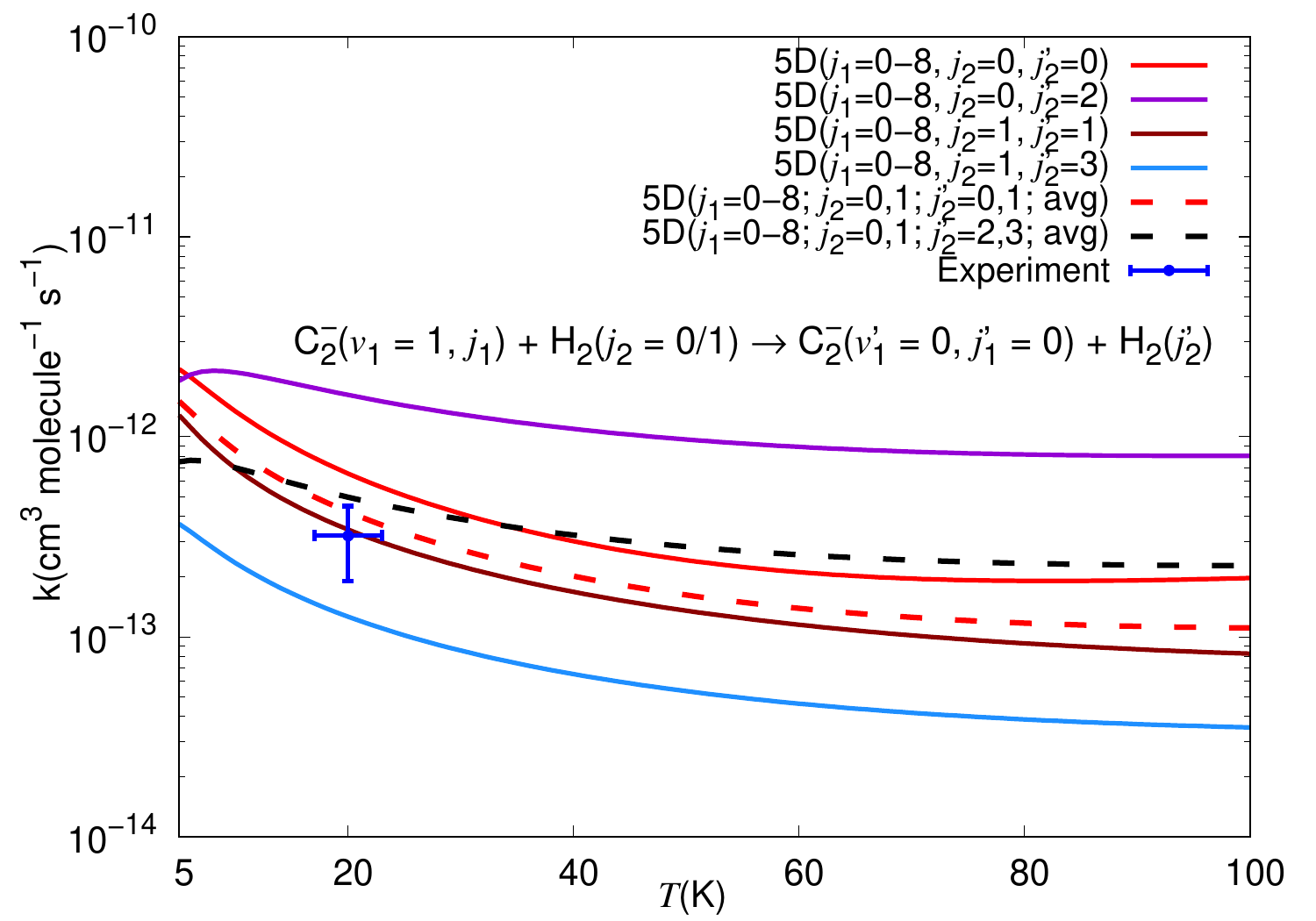}
\caption{Computed rate coefficients for vibrational de-excitation for the anionic partner with all the rotationally inelastic processes Boltzmann averaged at $T$ = 5-100 K. See the main text for details. The purple curve refers to 5D dynamics where the para-H$_2$ partner is rotationally excited from its ground state ($j_2$ = 0) to the $j_2'$ = 2  state, while the light blue curve refers to the ortho-H$_2$ partner getting excited from the $j_2$ = 1 state to the $j_2'$ = 3 state. Thus, the red  and brown curves only involve rotational states of the anion, while the purple and light blue curves describe the cases where either the para-H$_2$ or the ortho-H$_2$  partners is rotationally excited. All the initial states used to obtain these curves have been thermally averaged over $j_1$ = 0-to-8  initial states according to eq.(16) while the final rotational state of the anion was selected to be always the $j_1$ = 0 state. The dashed red and black curves report rate coefficient values obtained by averaging the dominant first three initial $j_1$ states with the experimental distribution at 20K  and further obtained by the weighted sum (3/4 + 1/4) of the results involving either the ortho- or the para-H$_2$ collision partner. In the case of the red dashes the H$_2$ partners are kept in their ground rotational states, while the dashed-black curve involves both ortho- and para- partners in their rotationally excited states.}
\label{fig: Figure16}
\end{figure}

There are two different aspects of these findings that require explanation : (i) why smaller energy transfers are yielding larger rate coefficients, and (ii) why limiting the final states to the pure vibrational cooling is bringing our calculations closer to the experimental data.

As for the first point, we already know that quantum dynamics yields in general larger probabilities for a final state to be reached whenever this event results in a smaller amount of energy being transferred. Hence, to produce a series of final anions which still retain some of their internal rotational energy is expected to happen with larger rate coefficients than when the amount of energy being transferred is larger, as when no internal energy is present in the final anion's rotational states.

However, explaining why this expected behavior brings our calculations closer to the experimental finding is still an open question, one which would imply that the energy redistribution in the experiments dramatically favors final states of the anion where no internal energy is left in their rotational states.

It is also interesting to note, as an aside, that the rate coefficients for  vibrational cooling occurring among the same levels of the present anion become dramatically smaller when He is used as a collision partner instead of H$_2$. For example, the cooling rate at 100 K, taken from our earlier work of reference \cite{20MaGiWe}, is about five orders of magnitude smaller than  the one we found here for the H$_2$ partner.

\section{Summary and Conclusions}
\label{sec:conc}
In the present work we have analysed in detail the quantum dynamics of an ion-neutral system, the ground electronic state of the C$_2^-$ interacting with the corresponding ground electronic state of the neutral H$_2$ molecule. More specifically, we have carried out detailed calculations for the dynamics of the vibrational cooling of the anion in collision with the hydrogen molecular partner. The general conditions of the overall system we have studied were  linked to low-temperature experiments done in our laboratory by injecting the two molecular partners in a cold ion trap, as described in more detail in one of our earlier papers, e.g. in \cite{23NWL}, while however also analysing in detail the complex dynamics of energy redistribution after collisions under low-T conditions.

By using new, accurate \textit{ab initio} calculations we have obtained a large grid of points involving 5 dimensions for the process at hand: rotations of both the anion and the neutral molecule and vibrations for the anion during the interaction with the H$_2$ partner. Such interaction PES was meant to improve on our earlier studies of the same system, when the H$_2$ presence was accounted for via an averaged potential for the neutral H$_2$ partner, thus using a 3D PES to treat the same dynamics (see further details in : \cite{23NWL}).

The main cooling process that we have considered here has been the vibrational transition from $\nu$ = 1 to $\nu$ = 0 of the anionic partner, together with a wide range of concurrent final populations of rotational states of the molecular ion in its ground vibrational state $\nu$ = 0. The H$_2$ partner has been considered to be a mixture of ortho- and para- states, as occurring in the experiment, and examples have been considered in the calculations where the neutral molecule ($p$-or $o$-H$_2$) was rotationally excited to either its $j_2$ = 2 or = 3 rotational states.

Several interesting features of the energy-transfer dynamics involving either the vibration or the rotations, or both of them, have been obtained from the present quantum dynamics, which is essentially an exact dynamics using the multichannel Coupled-Channel (CC) approach, as described in the earlier Section on the dynamics:

(i) both the state-to-state cross sections and the corresponding rate coefficients turn out to be larger when the final, vibrationally cooled, state of the anion was obtained into a variety of its  excited rotational states. In other words, we found that the purely vibrational energy transfer, yielding the largest energy gap between the involved states, was occurring with the smallest rate coefficients  while the concurrent rotational energy-transfer paths produced larger rates, up to fairly large transfers of rotation angular momenta. This type of behaviour was found to occur for both para- and ortho-H$_2$ as collision partners.

(ii) the efficiency of energy transfer processes by collision turned out to be larger for the para-H$_2$ partner, especially for transitions involving vibrational energy transfers. The present results  are thus  several orders of magnitude larger than in the case of inelastic rates computed when  He, or even Ar atoms are the collision partners: a result clearly reported by our earlier calculations of ref. \cite{20MaGiWe}.

(iii) our calculations also reveal that to employ the more realistic 5D version of the interaction forces, i.e. one where the H$_2$ molecular partner is considered also a rotating molecule, in comparison with the simpler 3D version of our earlier work in \cite{23NWL} where the hydrogen molecule was treated as a structureless partner, yields invariably larger inelastic rate coefficients (see comparison in figure 9). All the newly computed rate coefficients are seen, in fact, to be larger by varying amounts which are within about one order of magnitude of each other. In other words, the presence of a better description of the interacting systems causes the energy-transfer efficiency to the anionic partner by collision to uniformly increase for all processes.

(iv) When comparing the behavior of the neutral collision partners (i.e. ortho-H$_2$ and para-H$_2$), we found in our calculations, as described, for example, by the data in Figure 13, that to consider the ortho-H$_2$ as the collision partner turns out to consistently yield smaller collision rates for all the cases examined in the present study.

 The present calculations indicate that to obtain the final anion in its ground vibrational state, but also in a variety of excited rotational states, yields rate coefficients which are uniformly larger than when no rotational energy is being stored into the cooled-down molecular anion. 

To be more specific, the data reported by Figure 15 make a direct comparison with the only available experimental point, taken around 20K. They show that our essentially exact calculations are about one or two orders of magnitude larger than the measured datum when one factors into the modeling that the vibrational cooling via collisions occurs when the available internal energy  is  allowed to flow into a variety of excited rotational states of the final anions produced in the trap. On the other hand, only when the vibrationally cooled molecules are  allowed to reach their ground rotational state after the collisional process ( which is deemed to be a nonphysical option), hence producing anions with no internal  energy content,  we find that our modeling of the events now gets very close to the order of magnitude of the experimental findings.

Quantum inelastic dynamics already tell us that those processes, which involve the least amount of energy being transferred during a collision are occurring with larger probabilities, hence the differences in size between the top four curves and the dashed curve in Figure 15 are in accord with such a result.
On the other hand, to find that such decay option actually induces better agreement with the existing experiments is still an unexpected result, and one which   needs further investigation of  the energy-transfer dynamics in this specific system.  The new calculations presented in this work are indeed accurately detailing  the quantum dynamics of the investigated processes and able to clarify several aspects of the ro-vibrational inelastic collisions involving the two title molecules. Basically we have found that, once the correct 5D interaction is considered, and  exact quantum dynamics cross sections are obtained, the spin condition of the H$_2$ partner plays an important on the vibrational cooling efficiency. This is an unexpected result which is very useful in planning future cooling cycles of molecular anions.

\section{Acknowledgements}
All the numerical data pertaining to our parametric fitting of the computed PES and to the actual values of the computed cross
sections and rate coefficients are available in the Supplementary Information (SI) folder provided in the present paper. 
N.S. thanks the Indian National Science Academy for the award of INSA Distinguished Professorship. We are deeply grateful to Professor N. Balakrishnan and Professor R. Krems for generously providing us with a working version of the bimolecular rovibrational scattering code which we further specify in our references. We also Dr Katrin Erath-Dulitz for useful comments and suggestions on the manuscript in the making.
The research has been supported by the computer center WCSS (Wroclawskie Centrum Sieciowo-Superkomputerowe, Politechnika Wroclawska) through grant number KDM-408.
We acknowledge that the results of this research have been achieved using the DECI resource BEM2 based in Poland at WCSS with support from the PRACE aisbl.

\bibliography{c2m_Rg}

\widetext
\clearpage
\includegraphics[width=\textwidth]{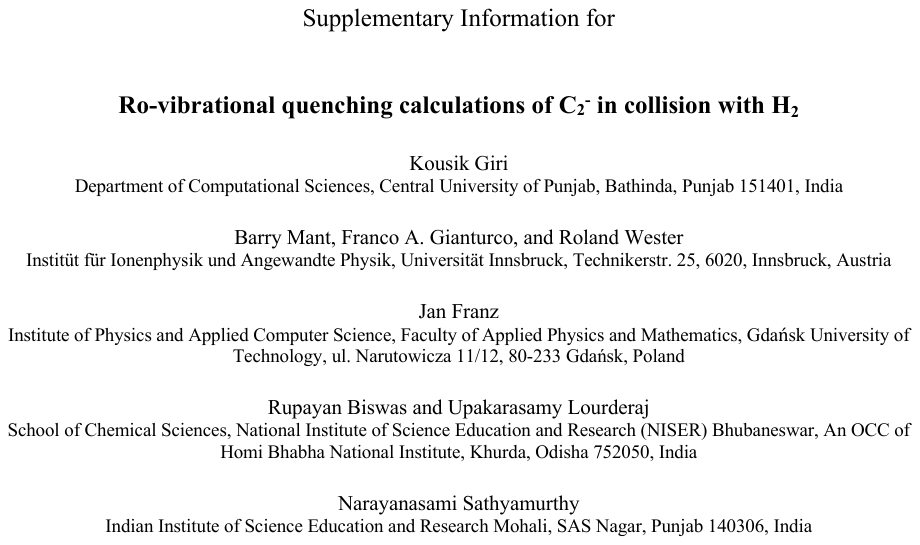}
\includegraphics[width=\textwidth]{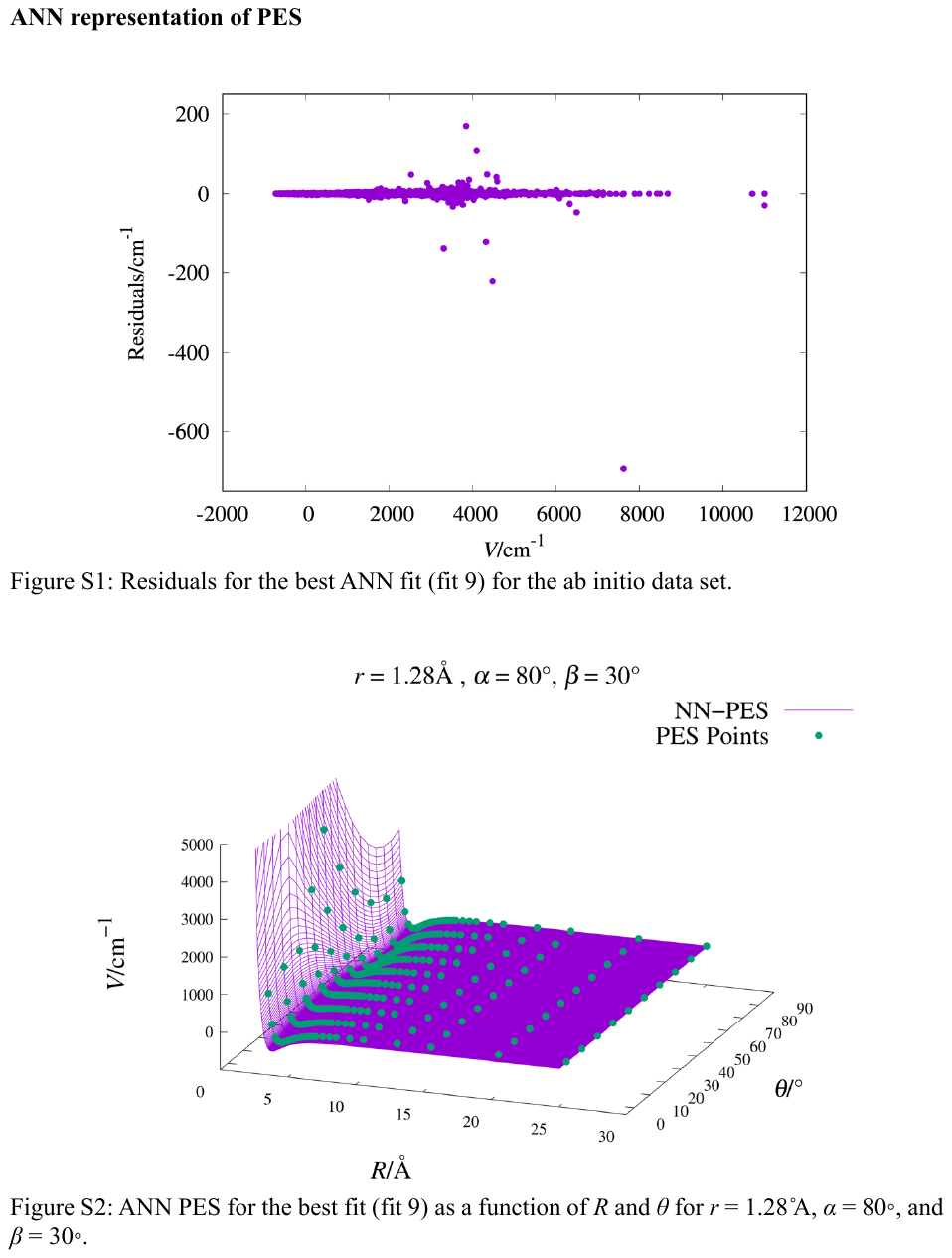}

\end{document}